%% file: ms.tex
\documentclass[hidelinks]{siamltex1213}
\pdfoutput=1
\usepackage[T1]{fontenc}
\usepackage{xcolor}
\usepackage{graphicx,multirow}
\usepackage{cancel}
\usepackage{amssymb,amsmath}
\usepackage{graphicx,amsmath,amsfonts,amssymb}
\usepackage{appendix}
\usepackage{xspace}
\usepackage{cite}
\usepackage{tensor}
\usepackage{algorithm}
\usepackage{algorithmicx}
\usepackage{algpseudocode}
\usepackage{standalone}
\usepackage{setspace}
\usepackage{array}
\usepackage{tabularx}
\usepackage{booktabs}
\usepackage{comment}
\usepackage{subcaption}
\usepackage{pgf}
\usepackage{pgfplots}
\usepackage{tikz}
 \usetikzlibrary{plotmarks,shapes,snakes,arrows,patterns,intersections,calc}
 \usetikzlibrary{external}
 \usepackage{pgfplotstable}
 \pgfplotsset{compat=1.11}
 \def\addlegendimage{\csname pgfplots@addlegendimage\endcsname}
 \usepgfplotslibrary{statistics}

\usepackage{tikzexternal}
\tikzsetexternalprefix{figures/}

\usepackage{hyperref}
\hypersetup{allcolors=black}

\input{preamble}
\title{Non-Linear Least-Squares Optimization of Rational Filters for the Solution of Interior Eigenvalue Problems
}%

\author{%
  Jan Winkelmann\thanks{Address: Aachen Institute for Advanced Study in Computational Engineering Science, RWTH Aachen University,
    Schinkelstr. 2, 52062--Aachen, Germany,  {\tt winkelmann@aices.rwth-aachen.de} }
  \and
  Edoardo Di Napoli\thanks{Address: J\"ulich Supercomputing Centre, Institute for Advanced
    Simulation, Forschungszentrum J\"ulich GmbH, Wilhelm-Johnen strasse,
    52425--J\"ulich, Germany.  {\tt e.di.napoli@fz-juelich.de}.}
}%
\begin{document}

\maketitle

\begin{abstract}
  Rational filter functions can be used to improve convergence of
  contour-based eigensolvers, a popular family of algorithms for the
  solution of the interior eigenvalue problem. We present a framework
  for the optimization of rational filters based on a non-convex
  weighted Least-Squares scheme. When used in combination with the
  FEAST library, our filters out-perform existing ones on a large
  and representative set of benchmark problems.  
  This work provides a detailed description of 
  : (1) a set up of the
  optimization process that exploits symmetries of the filter function
  for Hermitian eigenproblems, (2) a formulation of the gradient
  descent and Levenberg-Marquardt algorithms that exploits the
  symmetries, (3) a method to select the starting position for the
  optimization algorithms that reliably produces effective filters,
  (4) a constrained optimization scheme that produces filter functions
  with specific properties that may be beneficial to the performance
  of the eigensolver that employs them. 
 
\end{abstract}



\section{Introduction}
\input{introduction}

\section{Examples of \LSOP/ Filters}
\label{sec:results}
\input{results}

\section{Rational Filters for Hermitian Operators: A Non-Linear Least-Squares Formulation}
\label{sec:least-squares-formulation}
\input{ratfunc}

\input{optipole}

\section{Optimizing Rational Filters for Hermitian Operators}
\label{sec:optimizationMethods}
\input{numpole}

\input{concl}

\section*{Acknowledgments}
Financial support from the J\"ulich Aachen Research Alliance High Performance Computing and the
Deutsche Forschungsgemeinschaft (DFG) through grant GSC 111 is gratefully acknowledged.

\section*{Appendix}
\begin{appendices}
  \numberwithin{equation}{section}

  \input{appendix}

  \section{Obtaining benchmark problems from a matrix}
  \label{sec:app:ssst}
  \input{ssst}

  \section{List of filters}
  \label{sec:app:filterlist}
  \input{filters}

\end{appendices}

\bibliographystyle{siamplain}
\bibliography{./lsop}

\end{document}

%% file: preamble.tex
\def\myrefs#1#2{ 
{\bigskip \noindent
{\Large \bf #2}  
 \list {[\arabic{enumi}]}{\settowidth\labelwidth{[#1]}
 \leftmargin\labelwidth 
 \advance\leftmargin\labelsep
 \usecounter{enumi} }  
 \def\newblock{\hskip .11em plus .33em minus .07em}
 \sloppy\clubpenalty4000\widowpenalty4000
 \sfcode`\.=1000\relax}  }

\def\del{\partial} %
\def\diag{\mbox{Diag}\,}
\def\nref#1{(\ref{#1})}

\def\comb#1,#2,{ \left( {#1 \atop #2 } \right)  }%
\def\prodd#1,#2,#3,{ \prod_{\scriptstyle #1 \atop\scriptstyle #2 }^{#3} }%
\def\summ#1,#2,#3,{ \sum_{\scriptstyle #1 \atop\scriptstyle #2 }^{#3} }%
\def\RR{\mathbb{R}} 
\def\CC{\mathbb{C}}
\def\Hbb{\mathbb{H}}

\newtheorem{algor}{{\sc Algorithm}}[section]
\newtheorem{tabl}{Table}[section]

\def\betab{\begin{tabbing} 
xxxx\=xxxx\=xxx\=xx\=xx\=xx\=xx\=xx\=xx\=xx\=xx\=xx\=xx\= \kill} 
\def\entab{\end{tabbing}\vspace{-0.12in}}

\newcommand{\eq}[1]{\begin{equation}\label{#1}}
\newcommand{\en}{\end{equation}}

\newcommand{\beeq}[1]{\begin{equation}\label{#1}}
\newcommand{\eneq}{\end{equation}}

\graphicspath{{./FIGS/}}

\newcommand{\rotatevdash}{{\mathpalette\rotvdash\relax}}
\newcommand{\rotvdash}[2]{\rotatebox[origin=c]{270}{$#1\vDash$}}

\newcommand{\cnj}[1]{\ensuremath{\overline{#1}} \xspace}
\newcommand{\pty}[1]{\ensuremath{#1^\rotatevdash} \xspace}
\newcommand{\cvar}{\ensuremath{v} \xspace}

\def\CPPole{w}
\def\CPCoeff{\gamma}
\def\WeightFct{\mathfrak G}

\def\LSOP/{SLiSe}

\ifdefined\DEFCOLORS
\else
\definecolor{LSOPC}{RGB}{77,175,74}
\definecolor{GAUSSC}{RGB}{55,126,184}
\definecolor{ZOLOC}{RGB}{228,26,28}
\definecolor{LSOPC2}{RGB}{152,78,163}
\definecolor{LSOPC3}{RGB}{255,127,0}
\definecolor{XAXIS}{RGB}{100,100,100}
\newcommand*{\DEFCOLORS}{}%
\fi

\newtheorem{remark}{Remark}
\newtheorem{guideline}{Guideline}

%% file: introduction.tex
The last ten years have witnessed a proliferation of papers on methods
for the solution of the interior eigenvalue problem based on the use
of the resolvent operator
\cite{PhysRevB.79.115112,sakurai2003projection,Ikegami20101927,_feast_as_subsp_iterat_eigen}.
In practice, the spectrum of the eigenproblem can be ``sliced'' into
multiple subspaces leading to the solution of quasi-independent
sub-problems. Hence, these methods are considered to be part of the
larger class of spectrum-slicing eigensolvers. In many instances a
resolvent can been looked at as a rational function of the matrices
defining the eigenvalue problem. In this paper we conduct a detailed
investigation of how a carefully crafted rational function can improve
the efficacy of resolvent-based eigensolvers. To this end, we
initially adopt the conventional way to look at the resolvent as an
approximation of an ideal filter represented by the standard indicator
function. Then, the problem of finding a ``good'' rational function
can be cast as one of 1) finding a continuous filter that approximates
the discontinuous indicator function and, 2) improves the
effectiveness of the eigensolver by providing an extensive analysis of
the choice of parameters involved in the optimization process. We do
not aim at a filter that is just an optimal approximation of the
indicator function, but rather at a filter that is specifically
crafted for the class of eigensolver we are interested in. Our method
is based on a non-linear Least-Squares optimization approach, which
leads to a series of custom filters we termed \textbf{S}ymmetric
non-\textbf{L}inear Opt\textbf{i}mized Least-\textbf{S}quar\textbf{e}s
(\LSOP/), read ``slice''. Our work is a step towards problem-specific
filters that are flexible enough to be capable of exploiting cheaply
available information on the spectral structure of a given problem.
Such an approach can result in an eigensolver with better performance
and less parallel work-load imbalance. In addition, we provide an
efficient procedure that allows for the generation of such filters at
runtime.

\noindent The paper contains five major contributions:

\emph{Rich optimization framework.}  We illustrate an optimization
framework, via weighted Least-Squares, based on the $L_2$ function
norm instead of optimizing the rational function only at a number of
sample points via the $\ell_2$ norm (see \cite{VanBarel2016346}). The
defined residual level function, its gradients, and an approximated
Hessian are all based on the $L_2$ inner product. Such a setup allows
for the use of the Levenberg-Marquardt method to solve for the
non-linear Least-Squares problem. Flexibility is the landmark of the
developed framework: we present penalty parameters and box
constraints to obtain filters that, for example, are better suited for
Krylov-based linear system solvers.

\emph{Symmetries of the resolvent for Hermitian matrices.}  We expose
the intrinsic symmetries of the rational function approximating the
indicator function. Such symmetries are then exploited in the
implementation of the optimization process: filters which are
invariant under conjugation and parity transformations require fewer
degrees of freedom, which results in faster numerical executions.

\emph{Optimization of pole placement.}  Optimizing the rational
function's poles and coefficients requires the solution of a
non-convex problem. On the up side, this is a problem with a much
richer parameter space enabling more sophisticated solutions. On the
other hand, this is a much more complicated approach than previous
attempts \cite{saad16LeastSquares,VanBarel2016346}. If not carefully
constructed, naive realizations of the non-convex optimization may
often result in ineffective filters. There are many reasons
for ending up with an inferior filter: for instance the filter may be
asymmetric, be the outcome of an uncoverged optimization process, or
even have parameters hardly compatible with a spectrum-slicing
solver. In order to yield consistent results, we provide a selection
of initial parameters that consistently yield ``good'' filters.

\emph{Fire-and-forget optimization.}  We describe and discuss the
choices of parameters that might lead the optimization process astray,
ending up with ineffective filters. We propose a number of guidelines
that consistently yield filters that work well with solvers based on
the resolvent operator. In most cases no trial-and-error in the
parameter selection is required.
We argue at length on the principles that should guide the user in the
choice of Least-Squares weights, and constraints for the optimization.

\emph{Ready-to-use \LSOP/ filters.}  In order showcase the promise of
our optimization approach, we provide ready-to-use \LSOP/ filters. We
compare such filters with existing ones using a large representative
problem set. These \LSOP/ filters can be used as drop-in replacements
for the current state-of-the-art spectrum-slicing eigensolver
libraries such as FEAST.

\subsection{Subspace Iteration Method Accelerated via Spectral
  Projection}
\label{sec:1-2}
This section presents some mathematical background on subspace
iteration methods based on resolvents expressed as rational functions,
the main area of application that we will consider. Our focus is on
the Hermitian interior eigenvalue problem. Given a Hermitian matrix
$A = A^\dagger \in \CC^{n\times n}$ and a proper interval $[a,b]$,
with \([a,b] \cap [\lambda_1, \lambda_n] \neq \emptyset \),
we are interested in
finding the $\mathfrak m$ eigenpairs $(\lambda, v)$ inside $[a,b]$
resolving the secular equation:
\[
Av = \lambda v .
\]



An efficient subspace iteration method necessitates a projection
procedure that identifies a search subspace approximating the invariant
eigenspace corresponding to the eigenvalues lying in the $[a,b]$
interval. In the course of this paper we focus solely on methods that
achieve such projection through a rational function \(f(A)\)
of the matrix $A$. Such functions are also known as rational filters
although other filter forms, such as polynomial filters, do exist. A
rational filter of degree $(n-1,n)$ can be
expressed as a matrix function in partial fraction decomposition
\begin{equation} \label{eq:filter}
\sum_{i=1}^n \alpha_i (A - Iz_i)^{-1} = f(A)
\end{equation}
where $z_i \in \CC$ and $\alpha_i \in \CC$ are chosen in a way such
that the eigenvalues of $A$ inside $[a,b]$ are mapped to roughly one,
and the eigenvalues outside are mapped to roughly zero.  As a result,
the filter suppresses eigenvalues outside of $[a,b]$, which improves
convergence of the subspace iteration for the desired eigenvalues and
eigenvectors. In this context $f$, as a scalar function, can be seen
as a rational approximation to the indicator function $\chi_{[a,b]}$,
with unit value inside $[a,b]$ and zero everywhere else.  Filters are
usually generated for a search interval of $[-1,1]$ and then mapped to
$[a,b]$ via an appropriate transformation on the \(\alpha_i\)'s
and \(z_i\)'s.
Without loss of generality we always consider the search interval to
be $[-1,1]$. Forming $f(A)$ from Equation~\eqref{eq:filter} explicitly
involves the calculation of the matrix inverse. Using $f(A)$ within a
subspace iteration procedure requires only $X := f(A)Y$, which can be
rewritten as $p$ independent linear system solves:
\begin{equation} \label{eq:linsolve}
  (A - Iz_i) X = \alpha_i Y \qquad 1 \leq i \leq p
\end{equation}

If we assume that $f(A)$ outputs a good approximation to the search
subspace, a common form of subspace iteration is based on the use of
the Rayleigh-Ritz (RR) projection at each iteration. Provided that the
approximating subspace has a dimension $\mathfrak p$ equal or larger
than $\mathfrak m$, such a projection reduces the size of the
eigenproblem to be solved, and is guaranteed to output eigenpairs of
the reduced problem that are in a one-to-one correspondence with the
eigenpairs inside the search interval.

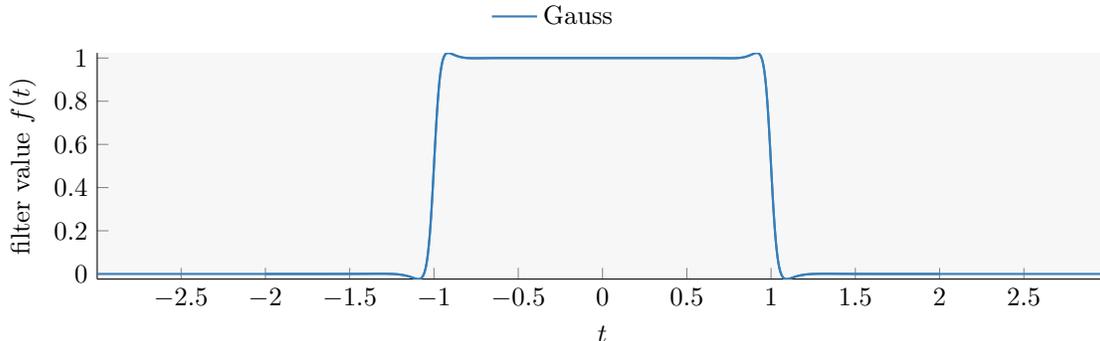
\begin{figure}
  \centering
  \ref{plotgauss}

  \begin{tikzpicture}
    \pgfplotstableread{./data/4_gausslike_plotdata.csv}{\fphiSD}
    \begin{axis}[
      enlargelimits=false,
      restrict x to domain=-3:3,
      axis y line*=left,
      axis x line*=bottom,
      xlabel={$t$}, ylabel={filter value $f(t)$},
      width=1.0\textwidth,
      height=0.2\textheight,
      axis background/.style={fill=black!3},
      legend columns=-1,
      legend entries={Gauss},
      legend to name=plotgauss,
      legend style={draw=none},
      ]
      \addplot+ [color=GAUSSC,mark=none, thick] table [x index={0},y expr=(\thisrowno{2})] {\fphiSD};
      \addplot+ [color=GAUSSC,mark=none, thick] table [x expr=-(\thisrowno{0}),y expr=(\thisrowno{2})] {\fphiSD};
    \end{axis}
  \end{tikzpicture}

  \vspace*{-0.5cm}
  \caption{Plot of the Gauss filter obtained by 16-points Gauss-Legendre quadrature of Equation~\eqref{eq:Cauchyfilter}, assuming \(t \in  \RR\).}
  \vspace*{-0.25cm}
  \label{fig:gaussPlot}
\end{figure}

One of the most well-known solvers for the interior eigenvalue
problem based on a rational resolvent followed by a
the RR projection is the FEAST library~\cite{PhysRevB.79.115112}.
FEAST derives the filter function via  $p$-point quadrature of the contour integral:
\begin{equation} \label{eq:Cauchyfilter}
\frac{1}{2\pi i} \int_\Gamma \frac{\text{d}t}{t - A} \approx \frac{1}{2\pi i} \sum_{i=1}^p w_i (A - Ix_i)^{-1} = f(A)
\end{equation}
where $\Gamma$ is a contour in the complex plane enclosing the search
interval $[a,b]$.  Quadrature along $\Gamma$ yields integration nodes
$x_i \in \CC$ and weights $w_i \in \CC$.  While any quadrature rule
can filter the eigenvalues to some extent, FEAST's defaults to the
Gauss-Legendre quadrature rules.
From Equation~\eqref{eq:Cauchyfilter} follows that numerical
integration of the contour integral is functionally equivalent to the
rational filter formulation of
Equation~\eqref{eq:filter}. Accordingly, we can interpret FEAST's
filter as a rational filter with \(x_i = z_i\)
and \(w_i = \alpha_i\);
we will call the resulting filter ``Gauss filter''. As we are
considering Hermitian eigenproblems, we can plot the Gauss filter by
considering $f(t)$ as a scalar function of $t \in \RR$.
Figure~\ref{fig:gaussPlot} plots the Gauss filter obtained by a
16-points quadrature on a circle-shaped contour that circumscribes
$[-1,1]$ symmetrically with respect to the real axis.  The resulting
function is even and real-valued on the entire real axis.

The convergence rate of the FEAST algorithm for a chosen filter \(f\)
is given by \cite{_feast_as_subsp_iterat_eigen}:
\begin{equation}
  \label{eq:convratio}
  \tau = \frac{ |f(\lambda_{\mathfrak p+1})|
  }{|f(\lambda_{\mathfrak m})|} 
\end{equation}
for a filter $f$ and an ordering of the eigenvalues such that
\[ 
|f(\lambda_1)| \geq |f(\lambda_2)| \geq ... \geq
|f(\lambda_{\mathfrak m})| \geq ... \geq |f(\lambda_{\mathfrak p})|
\geq ... \geq |f(\lambda_{\mathfrak p+1})| \geq ...  |f(\lambda_{n})|.  
\]
For most filters it is the case that \(f(1) = f(-1) = \frac{1}{2}\),
in which case we can extend the above equation with:
\mbox{\( |f(\lambda_{\mathfrak m})| \geq ... \geq \frac{1}{2} \geq
  ... \geq |f(\lambda_{\mathfrak p})| \)}.

The rest of this paper is structured as follows.  To showcase the
potential of our optimization method, Section~\ref{sec:results}
discusses example \LSOP/ filters.
Section~\ref{sec:least-squares-formulation} considers the
formulation of the non-linear Least-Squares problem.  We discuss the
use of symmetries of the filters, as well as the residual level
function and its gradients, which also make use of these symmetries.  In
Section~\ref{sec:optimizationMethods} we provide a full framework for
the generation of \LSOP/ filters.  Related work is presented in
Section~\ref{sec:relatedWork}, and we conclude in
Section~\ref{sec:conclusion}.


%% file: results.tex
In this section we provide a concise illustration of the main result
of the paper. Specifically, we present \LSOP/ filters that are meant
as replacements for existing rational functions used in
state-of-the-art filters. The methods by which these filters were
obtained are the topic of the following sections.
The filters used in this section, and the optimization parameters used to obtain them, are available in Appendix~\ref{sec:app:filterlist}.

First we discuss a replacement for the Gauss filter, as presented in
the previous section.  An improvement for this filter is particularly
usefull, since the Gauss filter and its variations are the default
choice for many contour solvers~\cite{EVSL,PhysRevB.79.115112}.  Next,
we discuss a filter that is used for problems where an endpoint of the
search interval coincides with, or is near, a large spectral cluster.
Such a scenario can occur within parallel solvers when multiple
adjacent search intervals are selected. 
All of the
filters we present in this section are primarily meant to show the
promise of \LSOP/ filters. 
For the sake of brevity we limit the examples to
filters with 16 poles and coefficients total. 
Naturally, the optimization methods presented in this work are
applicable to filters with an arbitrary number of poles.

\subsection{\texorpdfstring{$\gamma$-\LSOP/}{Gamma-\LSOP/}: A Replacement Candidate for the Gauss Filter}
\label{sec:org1c03c61}
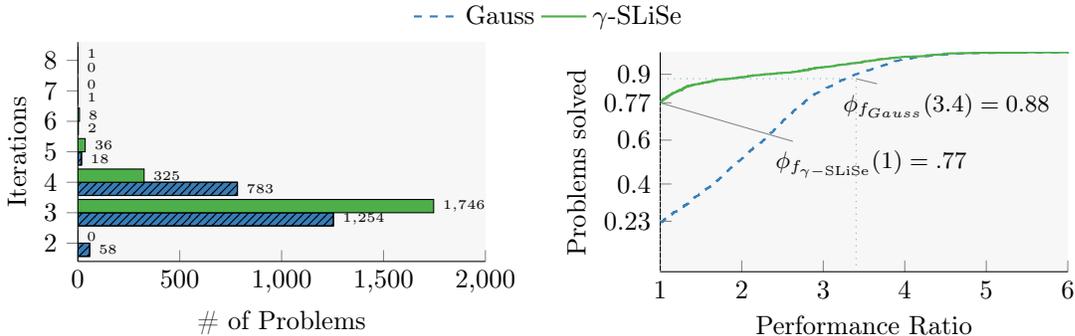
\begin{figure}
  \centering
  \ref{plotgammalsop}

  \begin{subfigure}[t]{.48\textwidth}
    \begin{tikzpicture}
      \pgfplotstableread{./data/si2_full_iterations.csv}{\fphiSD}
      \begin{axis}
        [ xbar=0pt
        , xmin = 0
        , xmax = 2000
        , bar width=5pt
        , axis y line*=left
        , axis x line*=bottom
        , axis background/.style={fill=black!3}
        , width=7cm
        , height=4.5cm
        , symbolic y coords={1,2,3,4,5,6,7,8}
        , ytick={1,2,3,4,5,6,7,8}
        , nodes near coords, nodes near coords align=horizontal
        , every node near coord/.append style={font=\tiny}
        , xlabel={\# of Problems}
        , ylabel={Iterations}
        ]
        \addplot [fill=GAUSSC, postaction={ pattern=north east lines } ] coordinates
	{
          (          58           ,2)
          (        1254           ,3)
          (         783           ,4)
          (          18           ,5)
          (           2           ,6)
          (           1           ,7)
          (           0           ,8)
        };

        \addplot [fill=LSOPC] coordinates
        {
          (           0           ,2)
          (        1746           ,3)
          (         325           ,4)
          (          36           ,5)
          (           8           ,6)
          (           0           ,7)
          (           1           ,8)
        };
      \end{axis}
    \end{tikzpicture}
    \caption{Number of FEAST iterations saved by $\gamma$-\LSOP/ over the Gauss filter.}
    \label{fig:lsop-gauss-si2}
  \end{subfigure}
  \begin{subfigure}[t]{.48\textwidth}
    \centering
    \begin{tikzpicture}
      \begin{axis}
        [ restrict x to domain=0:30
        , xmin=1
        , xmax=6
        , ymin=0
        , ymax=1
        , enlargelimits=false
        , axis y line*=left
        , axis x line*=bottom
        , width=7cm
        , height=4.5cm
        , axis background/.style={fill=black!3}
        , ytick={0.23,0.4,0.6,0.77,0.90}
        , xlabel={Performance Ratio}
        , ylabel={Problems solved}
        ]
        \addplot+[mark=none, thick,color=GAUSSC, dashed, name path=gaussgraph] table [x=x, y=y] {data/perfprof_Si2.mat_04_gamma.tex_Gauss};
        \addplot+[mark=none, thick, color=LSOPC, name path=lsopgraph] table [x=x, y=y] {data/perfprof_Si2.mat_04_gamma.tex_Gamma};

        \coordinate (o) at (1,0);

        \coordinate (inter1) at (1,0.77);
        \draw[dotted, color=LSOPC!70] (inter1|-o) -- (inter1) -- (inter1-|o);
        \node[coordinate, pin={[pin distance = 15 mm]340:{\small $\phi_{f_{\gamma-\text{\LSOP/}}}(1) = .77$}}]  at (inter1) {};

        \coordinate (inter2) at (3.4029,0.88);
        \draw[dotted, color=GAUSSC!70] (inter2|-o) -- (inter2) -- (inter2-|o);
        \node[coordinate, pin={[pin distance = 2 mm]340:{\hspace*{-6mm} \small $\phi_{f_{Gauss}}(3.4) = 0.88$}}]  at (inter2) {};

      \end{axis}
    \end{tikzpicture}
    \caption{Performance profiles of convergence rates for $\gamma$-\LSOP/ and Gauss filters}
    \label{fig:lsop-gauss-si2-pp-conv}
  \end{subfigure}
  \vspace*{-0.5cm}
  \caption{Comparisons of the Gauss filter with our replacement
    candidate, $\gamma$-\LSOP/, with $\mathfrak p = 1.5\mathfrak m$ on
    a testset of 2116 benchmark problems obtained from the "Si2"
    matrix. Both filters have 16 poles and are plotted in Figure~\ref{fig:gaussPlott}.}
  \label{fig:gaussPP}
  \vspace*{-0.25cm}
\end{figure}

In order to have a fair comparison between an existing filter and our
candidate as filter replacement, we need to establish a testing
environment. Since our focus is the use of rational filters within
subspace iteration methods, we use FEAST as our test environment for
comparing filters. A simple filter comparison could use the number of
FEAST iterations required to converge the entire subspace for a given
benchmark problem; the filter that requires fewer iterations is the
better one.  While rather basic, we show that this simple criterion
already provides useful insights.  Later in the paper we introduce a
more sophisticated evaluation criterion based on the convergence ratio
of the benchmark problem.

Once the comparison criterion has been selected, a natural way to
decide which filter is superior is to select many, representative
benchmark problems and see which filter performs better.  Comparing
filters on only a few, hand-selected, problems can
introduce a strong bias. In Appendix~\ref{sec:app:ssst} we propose a
method to obtain many benchmark problems from a given matrix which we
will use throughout this section. The resulting comparison is based on
a large number of benchmark problems and provides a good statistical
measure of
our filters' quality.

We construct a set of benchmark problems as follows: We fix the matrix
\(A\),
construct a large number of distinct search intervals \([a,b]\),
compute the exact number of eigenvalues $\mathfrak m$ each interval
contains, and then set \(\mathfrak p\)
to be a multiple of this number. The set of benchmark problems we will
use throughout this section were obtained by selecting 2116 search
intervals, each containing between 5\% and 20\% of the spectrum of
the Hermitian ``Si2'' matrix in the University of Florida matrix
collection \cite{Davis:2011:UFS:2049662.2049663}.
The Gauss and \(\gamma\)-\LSOP/
filter were used as filtering method in FEAST's version 3.0
\texttt{scsrevx} routine, a driver routine that uses a sparse direct
solver for the linear system solves\footnote{Feast was compiled with
  the Intel Compiler v16.0.2 and executed with a single thread; the
  target residual was adjusted to \(10^{-13}\)}.
For a problem with \(\mathfrak m\)
eigenpairs inside the search interval we select a size of the subspace
iteration of \(\mathfrak p=1.5\mathfrak m\),
the value that FEAST recommends for the Gauss filter.

Figure \ref{fig:lsop-gauss-si2} shows a histogram of the iterations
required for the 2116 benchmark problems by the \(\gamma\)-\LSOP/
and the Gauss filter. Independently of which 
filter is used the vast majority of problems of the set requires
either 3 or 4 FEAST iterations.  Fast convergence of FEAST's subspace
iteration is a known feature for the Gauss filter when \(\mathfrak p\)
is chosen large enough. When FEAST uses the \(\gamma\)-\LSOP/
filter, most benchmark problems require 3 iterations, with only some
problems requiring 4 or more iterations. In contrast, when the Gauss
filter is used, a larger number of problems require 4 iterations.
Summing the iterations for all of the benchmark problems the
\(\gamma\)-\LSOP/
filter requires 6774 iterations, while the Gauss filter requires 7119
iterations. Since every iteration requires 8 linear system solves and
additional overhead, such as the calculation of the residuals, saving
even a single iteration is a substantial performance improvement.

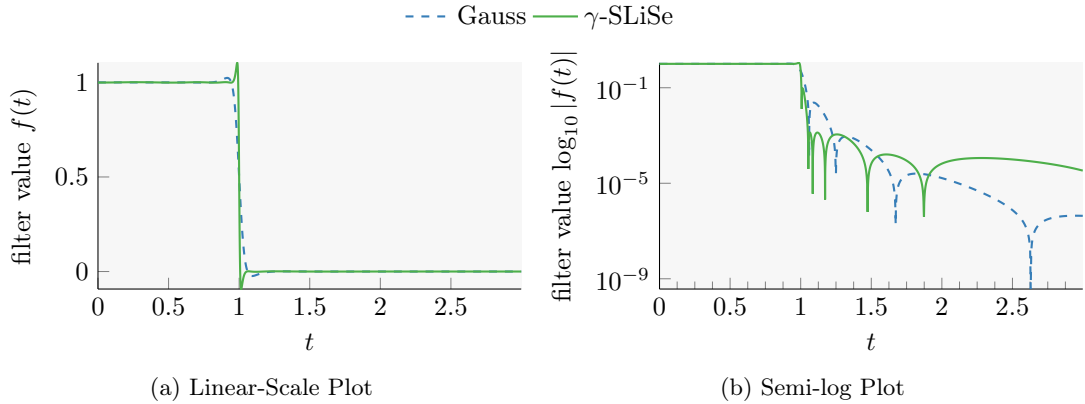
\begin{figure}
  \centering
  \ref{plotgammalsop}

  \begin{subfigure}[b]{.48\textwidth}
    \centering
    \begin{tikzpicture}
    \pgfplotstableread{./data/4_gausslike_plotdata.csv}{\fphiSD}
      \begin{axis}
        [ enlargelimits=false
        , restrict x to domain=0:3
        , axis y line*=left
        , axis x line*=bottom
        , xlabel={$t$}, ylabel={filter value $f(t)$}
        , width=1.0\textwidth
        , height=0.2\textheight
        , axis background/.style={fill=black!3}
        , legend columns=-1
        , legend entries={Gauss, $\gamma$-\LSOP/}
        , legend to name=plotgammalsop
        , legend style={draw=none}
        ]
        \addplot+ [color=GAUSSC, mark=none, thick, dashed] table [x index={0},y expr=(\thisrowno{2})] {\fphiSD};
        \addplot+ [color=LSOPC , mark=none, thick] table [x index={0},y expr=(\thisrowno{5})] {\fphiSD};
      \end{axis}
    \end{tikzpicture}
    \caption{Linear-Scale Plot}
    \label{fig:sd_filter:nonlog}
  \end{subfigure}
  \begin{subfigure}[b]{.48\textwidth}
    \centering
    \begin{tikzpicture}
    \pgfplotstableread{./data/4_gausslike_plotdata.csv}{\fphiSD}
      \begin{axis}
        [ width=1.0\textwidth
        , height=0.2\textheight
        , enlargelimits=false
        , restrict x to domain=0:3
        , axis y line*=left
        , axis x line*=bottom
        , xlabel={Performance Ratio}
        , ylabel={Problems solved}
        , xtick={0,0.5,...,3}
        , minor x tick num={3}
        , axis background/.style={fill=black!3}
        , ymode=log
        , xlabel={$t$}
        , ylabel={filter value $\log_{10} |f(t)|$}
        ]
        \addplot+ [color=GAUSSC,mark=none, thick, dashed] table [x index={0},y expr=abs(\thisrowno{2})] {\fphiSD};
        \addplot+ [color=LSOPC,mark=none, thick] table [x index={0},y expr=abs(\thisrowno{5})] {\fphiSD};
      \end{axis}
    \end{tikzpicture}
    \caption{Semi-log Plot}
    \label{fig:sd_filter:log}
  \end{subfigure}
  \vspace*{-0.5cm}
  \caption{Plots of the Gauss filter and our replacement candidate,
    $\gamma$-\LSOP/. Both filters have 16 poles and coefficients.}
  \label{fig:gaussPlott}
  \vspace*{-0.25cm}
\end{figure}

The increased performance of the \(\gamma\)-\LSOP/
filter comes with some drawbacks.  On the one hand, the Gauss filter
is more versatile than the \(\gamma\)-\LSOP/.
For instance, outside of the selected interval, the Gauss filter
decays very quickly to low values (see Figure~\ref{fig:sd_filter:log}),
which yields better convergence when increasing the number of vectors
in the subspace iterations \(\mathfrak p\)
to a larger multiple of \(\mathfrak m\).
The \(\gamma\)-\LSOP/
does not have this property, and so does worse when a very large
spectral gap is present. On the other hand, 
%
it is always possible to generate a
different \LSOP/ filter that performs well also for large spectral
gaps.

\subsubsection{Convergence Ratio as a Means of Comparison}


While Figure \ref{fig:lsop-gauss-si2} shows some promise as a tool for
comparing filters, the approach has a number of problems.  Subspace
iteration counts provide a very coarse look at the performance of a
filter, underlined by the fact that most benchmark problems require a
very similar number of iterations for both filters.  Further, the
evaluation requires the use of a linear system solver, and thus the
outcome is dependent on the algorithm and the parameters used (such as
the desired accuracy, the type of solver, etc.). A more reliable
comparison should be based on a more fine-grained metric which does
not depend on a specific solver, its specific parameters, and the
computational kernels used. One such metric is the rate of convergence
$\tau$, defined in Equation~\eqref{eq:convratio}, exhibited by the
filters on each problem of the benchmark set. For small to medium
sized problems, we can obtain all eigenvalues of the matrix and
calculate the convergence rate analytically. The rate of convergence
has a number of advantages over comparisons based on iteration counts.
First, it does not depend on a specific solver, and it is cheaper to
compute, as $\tau$ is a function of only the eigenvalues and not the
entire matrix. Then for testing purposes, we can compute all
eigenvalues up front, and then compute inexpensively \(\tau\)
for every benchmark problem.

Figure \ref{fig:lsop-gauss-si2-pp-conv}
shows the convergence rates for the same 2116 benchmarks problems as
in Figure \ref{fig:lsop-gauss-si2} in a form called \emph{performance
  profiles} \cite{dolan02_bench_optim_softw_with_perfor_profil}.  For
a filter \(f\),
and a given point \(x\)
on the abcissa, the corresponding value \(\phi_f(x)\)
of the graph indicates that for \(100\cdot\phi_f(x)\)
percent of the benchmark problems the filter \(f\)
is at most a factor of \(x\)
worse than the fastest of all methods in question. For example,
\(\phi_{f_{Gauss}}(3.4) \approx 0.88\)
indicates that for 88\% of the benchmark problems the Gauss filter
yields either the best convergence rate or a rate that is within a
factor of \(3.4\)
from the best.  So the performance profiles not only report how often
a filter performs best, but also how badly the filter performs when it
is not the best.  We will use performance profiles multiples times in
this section to compare the filters.
From the value of \(\phi_{f_{\gamma\text{-\LSOP/}}}(1)\)
in Figure \ref{fig:lsop-gauss-si2-pp-conv} we can infer that
\(\gamma\)-\LSOP/
achieves the best convergence rate for 77\% of the benchmark problems,
while the Gauss filter does best on the remaining 23\% the problems.
Moreover, the Gauss filter achieves a convergence rate that is worse
than 3.4 times that of the \(\gamma\)-\LSOP/
filter for 12\% of the problems
\(\phi_{f_{\text{Gauss}}}(3.4) \approx 0.88\). In conclusion,
the \LSOP/ filter performs better than the Gauss filter in both
metrics, iterations counts and convergence rate.  However, the
performance profile gives a much more detailed comparison.


Figure \ref{fig:sd_filter:nonlog} plots both filter functions; only
the positive part of the abscissa is shown as the functions are even.
Figure \ref{fig:sd_filter:log} shows the absolute value of the same
filter functions as a logplot.  The \(\gamma\)-\LSOP/
filter performs better than the Gauss filter due to a smaller absolute
value between 1.1 and 1.6 than the Gauss filter.  A smaller absolute
value outside the search interval---specifically for
\(|f(\lambda_{\mathfrak p+1})|\)---improves
the convergence rate \(\tau\).
We set \(\mathfrak p\)
to the recommended value of \(1.5\mathfrak m\).
A roughly equidistant spacing of the eigenvalues around the canonical
search interval \([-1,1]\)
yields \(\lambda_{\mathfrak p+1} \approx \pm 1.25\).
The \(\gamma\)-\LSOP/
filter out-performs the state-of-the-art Gauss filter for
\(\lambda_{\mathfrak p+1} \in [1.1,1.6]\).
If either end of the search interval is near a large spectral cluster
then \(\lambda_{\mathfrak p+1} \ll 1.1\),
and neither of the two filters will do particularly well.  In such a
case a larger \(\mathfrak p\)
is chosen, or Elliptic filters are used.  We present the Elliptic
filter---and a \LSOP/ filter meant to replace it---in the next
subsection.

\subsection{\texorpdfstring{$\eta$-\LSOP/}{Eta-\LSOP/}: A Replacement Candidate for the Elliptic Filter}
\label{sec:results:eta}

\begin{figure}
  \centering
  \ref{plotetalsop}

  \begin{subfigure}[b]{.48\textwidth}
    \centering
    \begin{tikzpicture}
      \begin{axis}
        [ width=.98\textwidth
        , height=0.2\textheight
        , enlargelimits=false
        , ymin=0
        , ymax=1
        , legend columns=-1
        , legend entries={Gauss,$\eta$-\LSOP/,Elliptic}
        , legend to name=plotetalsop
        , legend style={draw=none}
        , axis background/.style={fill=black!3}
        , ytick={0.1,0.3,0.5,0.7,0.9}
        , xtick={1,2,5,10,20,70}
        , minor xtick={1,...,10,20,30,40,50,60}
        , log ticks with fixed point
        , extra x tick style={/pgfplots/major tick length=.2cm,/pgfplots/tick style={line width=1.5pt}}
        , xlabel={Performance Ratio (log)}
        , ylabel={Problems solved}
        , axis y line*=left
        , axis x line*=bottom
        , xmode=log
        , y tick label style={/pgf/number format/.cd, fixed, precision=2, /tikz/.cd }
        ]

       \addplot+[mark=none, thick, color=GAUSSC, dashed, name path=gaussgraph] table [x=x, y=y] {data/eta.tex_Gauss};
       \addplot+[mark=none, thick, color=LSOPC, name path=lsopgraph] table [x=x, y=y] {data/eta.tex_eta};
       \addplot+[mark=none, thick, color=ZOLOC, dashdotted, name path=zolograph] table [x=x, y=y] {data/eta.tex_Elliptic};

      \end{axis}
    \end{tikzpicture}
    \caption{Semi-log performance profile of the convergence rate.}
    \label{fig:eta_4:nozoom}
  \end{subfigure}
  \begin{subfigure}[b]{.48\textwidth}
    \centering
    \begin{tikzpicture}
      \begin{axis}
        [ width=.98\textwidth
        , height=0.2\textheight
        , enlargelimits=false
        , ymin=0
        , ymax=1
        , xmin=1
        , xmax=3
        , axis background/.style={fill=black!3}
        , ytick={0.1,0.3,0.5,0.7,0.9}
        , xtick={1,1.38,2,2.35,3}
        , xlabel={Performance Ratio}
        , ylabel={Problems solved}
        , axis y line*=left
        , axis x line*=bottom
        , y tick label style={/pgf/number format/.cd, fixed, precision=2, /tikz/.cd  }
        ]

       \addplot+[mark=none, thick, color=GAUSSC, dashed, name path=gaussgraph] table [x=x, y=y] {data/eta.tex_Gauss};
      \addplot+[mark=none, thick,  color=LSOPC, name path=lsopgraph] table [x=x, y=y] {data/eta.tex_eta};
      \addplot+[mark=none, thick, color=ZOLOC, dashdotted, name path=zolograph] table [x=x, y=y] {data/eta.tex_Elliptic};

        \coordinate (o) at (1,0);

        \coordinate (o) at (1,0);

        \coordinate (inter1) at (2.35,1.0000);
        \draw[dotted, color=ZOLOC!70] (inter1|-o) -- (inter1) -- (inter1-|o);
        \node[coordinate, pin={[pin distance = 19 mm]270:{\hspace*{-5mm}\small $\phi_{f_{\text{Elliptic}}}(2.35)=.99$}}]  at (inter1) {};
        \coordinate (inter2) at (1.38,1.0000);
        \draw[dotted, color=LSOPC!70] (inter2|-o) -- (inter2) -- (inter2-|o);
        \node[coordinate, pin={[pin distance = 10 mm]300:{\small $\phi_{f_{\eta-\text{\LSOP/}}}(1.38)=.99$}}]  at (inter2) {};
      \end{axis}
    \end{tikzpicture}
    \caption{The same performance profile zoomed into \([1,3]\) with a linear axis}
    \label{fig:eta_4:zoom}
  \end{subfigure}
  \vspace*{-0.5cm}
  \caption{Semi-log performance profile of the convergence rate for
    the Gauss, $\eta$-\LSOP/, and Elliptic filters. $\eta$-\LSOP/ is
    meant to replace the Elliptic filter. All filters have 16 poles and $\mathfrak p = \mathfrak m$.}
  \label{fig:eta_4}
  \vspace*{-0.25cm}
\end{figure}
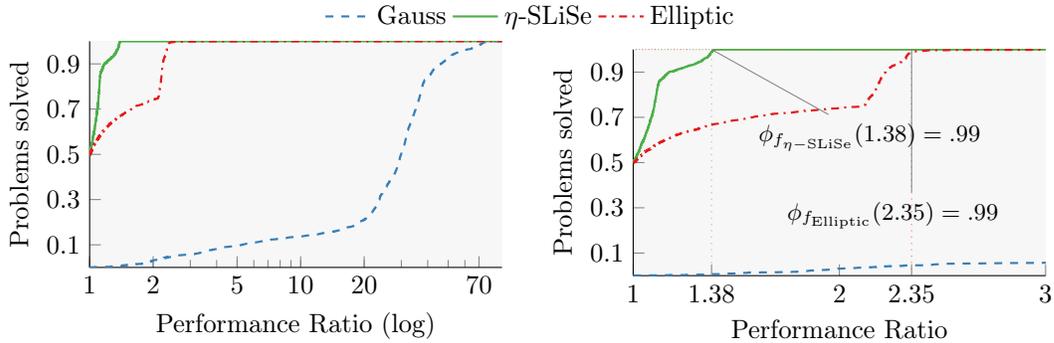

The Elliptic filter, also called Cauer or Zolotarev filter, has been
proposed for use in the context of the interior eigenvalue problem in
a number of publications
\cite{weko_67773_1,weko_69671_1,weko_28810_1,weko_28690_1,weko_18206_1,2014arXiv1407.8078G}\footnote{Technically
  the Elliptic filter is a class of filters depending on a number of
  paramters, we consider the specific filter discussed in
  \cite{2014arXiv1407.8078G}}. This filter is used specifically when large
spectral clusters are at or near the endpoints of a search interval.
In these cases the Gauss filter exhibits
very slow convergence unless \(\mathfrak p\)
is chosen much larger than \(1.5\mathfrak m\).
We propose \(\eta\)-\LSOP/ as a replacement for the Elliptic filter
currently used in contour-based subspace iteration methods.

The Elliptic filter is an optimal filter, in the sense that the filter
function is the best uniform rational approximation of the indicator
function.  Such a filter is worst-case optimal with respect to the
convergence rate of Equation~\eqref{eq:convratio}, for a subspace size
of \(\mathfrak p=\mathfrak m\).  In simple terms, the Elliptic filter
works better than the Gauss filter in the presence of spectral
clusters close to the interval boundary because it drops from 1 to 0
more quickly.  The \(\eta\)-\LSOP/ filter trades off a slightly larger
absolute values inside \([1,1.01]\) for smaller absolute values in
\([1.01,\infty]\) (\([-1.01,1]\) and \([-\infty, 1.01]\), respectively).
In Section~\ref{sec:additional} we elaborate on the choice of optimization parameters that yield the \(\eta\)-\LSOP/ filter.

Figure \ref{fig:eta_4} shows a performance profile of the
convergence rate for the Gauss, \(\eta\)-\LSOP/,
and Elliptic filters.  Figure \ref{fig:eta_4:nozoom} is a semi-log
plot of the performance profile, while Figure
\ref{fig:eta_4:zoom} is a linear-scale plot, zoomed into the
interval \([1,3]\).
The comparison between the two filters needs to be adjusted to represent
the specific use case of spectral clusters near the endpoints. We
reuse the 2116 benchmark problems from the ``Si2'' matrix from the
previous comparison, however, instead of target subspace size of
\(\mathfrak p=1.5\mathfrak m\),
we choose \(\mathfrak p=\mathfrak m\).
Thus, the  eigenvalue of largest absolute function value outside the search interval determines
convergence for all benchmark problems.  For large spectral clusters
\(|f(\lambda_{\mathfrak p+1})|\)
is likely to be close to \(|f(\lambda_{\mathfrak m})|\),
so this metric serves as an estimator of the filter's behaviour with
spectral clusters near the endpoints of the search interval.

The figure indicates that, for our representative set of benchmark
problems, the \(\eta\)-\LSOP/
filter displays better convergence rates than the Elliptic filter. The
Elliptic filter achieves the best convergence ratio for half of the
benchmark problems, and \(\eta\)-\LSOP/
does it for the other half.  Further, the figure indicates that for
99\% of the problems, the Elliptic filter exhibits a convergence ratio
within a larger factor from the convergenc ratio of \(\eta\)-\LSOP/
with $\phi_{f_{\text{Elliptic}}}(2.35)=.99$, and
$\phi_{f_{\eta-\text{\LSOP/}}}(1.38)=.99$ respectively.
That is, for 99\% of the problems \(\eta\)-\LSOP/
performs within a factor of 1.38 of the convergence ratio of the
Elliptic filter.  Simply put, \(\eta\)-\LSOP/
performs better than the Elliptic filter for half of the problems, and
for the other half it performs very similar to the Ellitical filter.
Conversely, the Elliptical filter exhibits convergence rates for some
problems that are twice as bad as the convergence rate of
\(\eta\)-\LSOP/.
The Gauss filter exhibits the worst rates of convergence by far, but
this is to be expected as the Gauss filter is not meant for this
use-case.





%% file: ratfunc.tex
Before discussing the actual optimization, we need to reformulate the
mathematical problem. Rational filters can be written as a sum of
simple rational functions $\phi_i$
\begin{equation}
\label{eq:ratfunapp}
f(t,z,\alpha) = \sum^n_{i=1} \phi_i \equiv
 \sum^n_{i=1}\frac{\alpha_i}{(t-z_i)}
\end{equation}
with $z_i, \alpha_i \in \CC$.  We restrict ourselves to the case where
the \(z_i\)
and the \(\alpha_i\)
are pair-wise distinct, and the \(z_i\)
have non-zero real and imaginary parts. Our initial goal is to set the
stage for the formulation of a reliable method dictating the choice for
the poles $z_i$ and the coefficients $\alpha_i$ such that $f(t)$
constitutes an effective and flexible approximation to the ideal
filter \(h(t)\).
In the Hermitian case such an ideal filter corresponds to the
indicator function defined as
\begin{align}
\label{eq:indicator}
h(t) = \chi_{[-1, 1]} = \left\{\begin{array}{l l}
1  & \quad \textrm{if}\ -1 \leq t \leq 1 \\
0  & \quad \textrm{otherwise}\\
\end{array}\right. .
\end{align}

As $h(t)$ is discontinuous 
we have to specify in what sense a given filter $f(t,z,\alpha)$ is
close to the target function $h(t)$. To this end, we focus on the
squared difference of $h(t) - f(t,z,\alpha)$ with respect to the
\(L_2\)
norm which is equivalent to defining a Least-Squares error function,
or \emph{residual level function}
\begin{equation}
\label{eq:reslev}
F(z,\alpha) = \| h - f(z,\alpha) \|_2^2 .
\end{equation}
An optimal filter, in this sense, in the minimizer of the
residual level function
\begin{align}\label{eq:optproblem}
  \min_{\alpha_i, z_i \; 1 \leq i \leq n} \;  \int_{-\infty}^{\infty}
  \WeightFct(t) \left| \, h(t) - \sum_{i=1}^n
  \frac{\alpha_i}{(t-z_i)} \, \right|^2  dt ,
\end{align}
resulting in a weighted Least-Squares optimization problem with the
weight function \(\WeightFct(t)\).
In order to simplify the optimization process we impose the following
restrictions on the weight function: 1) $\WeightFct(t)$ is piece-wise
constant with bounded interval endpoints, 2) the number of piece-wise
constant parts is finite, 3) the weight function is even:
$\WeightFct(t) = \WeightFct(-t)$.


In Section~\ref{sec:2-1} we illustrate how to
exploit the symmetries of the indicator function $h(t)$ in the
Hermitian case. The rational function $f(t)$ approximating \(h(t)\)
can be constructed so as to explicitly satisfy these symmetries, which
limits the degrees of freedom used in the optimization
problem. Section~\ref{sec:3-1} contains a formulation of the
Least-Squares residual level function, while its gradients are
introduced in Section~\ref{sec:gradients}.

\subsection{Discrete Symmetries of the Rational Approximant}
\label{sec:2-1}
We wish to construct a rational function $f(t,z,\alpha)$ that explicitly preserves the
symmetries of the target function $h(t)$. The target function is invariant under two
symmetry transformations, namely {\em complex conjugation}, and {\em
  reflection}. The first symmetry seems trivial since $h$ is
real-valued but $f$ in general is not. The second symmetry states that
$h$ is even.  For the rational function $f$ to be a good approximation
of $h$ it should satisfy both symmetries
\begin{equation}
\label{eq:invsym}
\begin{array}{l l r}
1. & \text{Complex Conjugation (C):}\qquad & \cnj{f}(t,z,\alpha) =
     f(t,z,\alpha) \\
2. & \text{Reflection or Parity (P):} \qquad & f(-t,z,\alpha) = f(t,z,\alpha)
\end{array}
\end{equation}

Requiring the function $f$ to satisfy the C symmetry implies that half
of the rational monomials $\phi_i$s are conjugate of each other. In
other words, half of the poles $z_i$s (and corresponding $\alpha_i$s) are in
the upper-half of $\CC$ (indicated as $\Hbb^+$) with the other half in
the lower-half ($\Hbb^-$). Then, without loss of generality we can
enumerate the $\phi_i$ such that the first $p={n \over 2}$ have poles
$z_i$ in $\Hbb^+$: \mbox{\(\frac{\alpha_r}{t-z_r} = \phi_r = \cnj{\phi_{r+p}} \)} with \(z_r \in \Hbb^+\) for \( 1 \leq r \leq p\).
This conclusion enables us to rewrite the sum over monomials $\phi_i$
as a sum over half the index range \(1 \leq j \leq p={n \over 2}\)
\begin{equation}
\label{eq:ratfun1sym}
f(t,\cvar,\beta) = \sum^p_{j=1} \left(\psi_j + \cnj{\psi}_j \right)  \equiv
 \sum^p_{j=1} \left[\frac{\beta_j}{(t-\cvar_j)} +
   \frac{\cnj{\beta}_j}{(t-\cnj{\cvar}_j)} \right],
\end{equation}
where we relabeled the poles $z_i$s, the coefficients $\alpha_i$s, and
the monomials $\phi_i$s to make the symmetry explicit.
\begin{equation}
\label{eq:relab}
\begin{array}{r l r l r l}
z_j = & \cvar_j \quad & \alpha_j = & \beta_j \quad & \phi_j = & \psi_j \\
z_{j+p} = & \cnj{\cvar}_j \quad & \alpha_{j+p} = & \cnj{\beta}_j \quad &
\phi_{j+p} = & \cnj{\psi}_j
\end{array} .
\end{equation}
Notice that satisfying the C symmetry forces the association of conjugate
coefficients $\cnj{\beta}_j$ with conjugate poles $\cnj{\cvar}_j$.


Imposing the P symmetry explicitly is a bit more tricky.  It can be
visualized as a symmetry for the complex numbers $\cvar_j, \beta_j$
and their conjugates, living in the complex plane.  Consider the poles
\(\cvar_k\)
in the right (R) upper-half of the complex plane $\Hbb^{+R}$, that is
those \(\cvar_k\)
with \(\Re e(\cvar_k) > 0\)
and \(\Im m(\cvar_k) > 0\).
Let the number of \(\cvar_k\)
in $\Hbb^{+R}$ be \(q\).
Applying the P symmetry to the monomials \(\psi_k\)
with poles \(\cvar_k\) in $\Hbb^{+R}$ yields:
\[
\psi_k(t,\cvar, \beta) = \frac{\beta_k}{(t-\cvar_k)} \longrightarrow  -\frac{\beta_k}{(t+\cvar_k)} = \psi_k(t,-\cvar,-\beta)
\]
Consequently the reflection operation maps a rational monomial with a
pole in $\Hbb^{+R}$ to a monomial with a pole in $\Hbb^{-L}$.  For $f$
to be invariant under reflection, the $q$ $\psi_j$ monomials with poles
in $\Hbb^{+R}$ ``must'' map to the $\cnj{\psi}_j$ monomials with poles
in $\Hbb^{-L}$.  By complex conjugation, these same $q$ $\psi_j$
monomials map to $\cnj{\psi}_j$ monomials in $\Hbb^{-R}$, consequently
$p = 2 q$.  Now we can enumerate the monomials such that the first
half of $\psi_j$ are in $\Hbb^{+R}$, the second half of $\psi_j$ are
in the $\Hbb^{+L}$, the first half of $\cnj{\psi}_j$ are in
$\Hbb^{-R}$ and the last $q$ $\cnj{\psi}_j$ are in $\Hbb^{-L}$.
Invariance under reflection implies then that
$\psi_k(t,-\cvar,-\beta) =
\cnj{\psi}_{k+q}(t,\cnj{\cvar},\cnj{\beta})$
for \(1 \leq k \leq q \).
The same reasoning can be repeated for monomials $\psi_k$ in
$\Hbb^{+L}$.  Finally, we can express $f$ by summing over a reduced
range $1 \leq k \leq q$ as
\begin{equation}
\label{eq:ratfun2sym}
f(t,\CPPole,\CPCoeff ) = \sum^q_{k=1} \left( \chi_k + \cnj{\chi}_k +
  \pty{\chi}_k + \pty{\cnj{\chi}}_k \right) = \sum^q_{k=1}
\left[ \frac{\CPCoeff_k}{t - \CPPole_k} + \frac{\cnj{\CPCoeff}_k}{t - \cnj{\CPPole}_k}
  - \frac{\CPCoeff_k}{t + \CPPole_k} - \frac{\cnj{\CPCoeff}_k}{t + \cnj{\CPPole}_k} \right].
\end{equation}     
Once again, we have relabeled poles, coefficients, and monomials
so as to make explicit the symmetry indicated by the $\rotatevdash$ symbol
\begin{equation}
\label{eq:relab1}
\begin{array}{r l r l r l}
\cvar_k = & \CPPole_k \qquad & \beta_k = & \CPCoeff_k \qquad & \psi_k = & \chi_k \\
\cvar_{k+q} = & \pty{\cnj \CPPole}_k = -\cnj{\CPPole}_k \qquad & \beta_{k+q} = & 
\pty{\cnj \CPCoeff}_k = -\cnj{\CPCoeff}_k \qquad & \psi_{k+q} = & \pty{\cnj{\chi}}_k \\
\cnj{\cvar}_k = & \cnj{\CPPole}_k \qquad & \cnj{\beta}_k = & \cnj{\CPCoeff}_k \qquad 
& \cnj{\psi}_k = & \cnj{\chi}_k \\
\cnj{\cvar}_{k+q} = & \pty \CPPole_k = - \CPPole_k \qquad & \cnj{\beta}_{k+q} = 
& \pty \CPCoeff_k = - \CPCoeff_k \qquad & \cnj{\psi}_{k+q} = & \pty{\chi}_k 
\end{array} .
\end{equation}
We would have reached the same result if we started to
require invariance under the symmetries in reverse
order\footnote{Physicists refer to the combination of these two
  discrete symmetries as CP invariance. A necessary condition for its
  existence is that the operators generating the transformations must
  commute.}.

%% file: optipole.tex
\subsection{Residual Level Function}
\label{sec:3-1}
Let us expand Equation~\eqref{eq:reslev} by temporarily disregarding the
discrete symmetries of the rational function and expressing $f$ as in
Equation~\eqref{eq:ratfunapp}
\begin{align*}
  F(z,\alpha)
&=\| h - \sum_i \alpha_i \phi_i \|_2^2 \\
&=  < h - \sum_i \alpha_i \phi_i , h - \sum_i   \alpha_j   \phi_j >\\
&=  < h,h>  - <h , \sum_j   \alpha_j   \phi_j> - 
< \sum_i \alpha_i \phi_i, h> 
 + < \sum_i \alpha_i \phi_i \ , \ \sum_j   \alpha_j   \phi_j >\\
&=  < h,h>  -  2  \Re e [ \sum_i <\phi_i , h> \alpha_i ]
 + \sum_i \sum_j \cnj \alpha_i \alpha_j < \phi_i \ ,   \phi_j >.
\end{align*}
The residual level function is expressible as
\begin{equation}
\label{eq:reslev1}
F(z, \alpha) = 
\alpha^\dagger G \alpha - 2 \Re e [ \eta^\dagger \alpha  ] + \|h\|^2
\end{equation}
where the $x^\dagger \equiv \cnj x^\top$ indicates complex conjugation plus transposition
(Hermitian conjugation) and
\[
G_{ij} = <\phi_i,   \phi_j> 
 ,\quad \eta_i =  <\phi_i, h> . 
\] 
The inner products $< \cdot, \cdot >$ are defined through a weight
function $\WeightFct (t)$
\begin{equation}
\label{eq:innpro}
<\phi_i,   \phi_j> 
= \int_{-\infty }^{+\infty} \WeightFct (t) \cnj{\phi_i}(t) \phi_j(t) dt 
= \int_{-\infty }^{+\infty} \frac{\WeightFct (t)}{(t-\cnj z_i) (t- z_j)} \ dt ,
\end{equation}
and
\[ 
\eta_i = <\phi_i, h> = 
 \int_{-\infty }^{+\infty} \frac{\WeightFct (t) h(t) }{t- \bar z_i} \ dt 
\qquad   \|h\|^2 = 
 \int_{-\infty }^{+\infty} \WeightFct (t) h(t)^2 \ dt.
\]



We could have started directly with the CP invariant formulation
from Equation~\eqref{eq:ratfun2sym} of the rational function $f$. We preferred this
approach in order to show how requiring each symmetry to be satisfied
has direct consequences on the structure of the matrix $G$ and
vector $\eta$. Recall from Equation~\eqref{eq:relab} how the $z_i$s and
$\alpha_i$s were mapped to $\cvar_j$s and $\beta_j$s so as to make explicit the C symmetry.
Due to such map, the residual level function has the following 
block structure:
\begin{equation}
\label{eq:blkreslev}
F =
\begin{pmatrix}
\beta^\dagger & \beta^\top \end{pmatrix}
\begin{pmatrix}
A & B \\ \cnj B & \cnj A \end{pmatrix}
\begin{pmatrix}
\beta \\ \cnj \beta \end{pmatrix}
- 2 \Re e \left[
\begin{pmatrix}
\zeta^\dagger & \zeta^\top \end{pmatrix}
\begin{pmatrix}
\beta \\  \cnj \beta  \end{pmatrix}
\right] + \|h\|^2
\end{equation}
with 
\begin{align*}
A_{i,j} = \int_{-\infty }^{+\infty} \frac{\WeightFct (t)}{(t- \cnj \cvar_i)(t - \cvar_j)} \ dt
& \quad \zeta_i = \int_{-\infty }^{+\infty} \frac{\WeightFct (t) h(t) }{t- \cnj \cvar_i} \ dt \\
B_{\underset{i\neq j}{i,j}} =  \int_{-\infty }^{+\infty}
  \frac{\WeightFct (t)}{(t- \cnj \cvar_i)(t - \cnj \cvar_j)} \ dt
& \quad B_{i,i} =  \int_{-\infty }^{+\infty} \frac{\WeightFct (t)}{(t- \cnj
  \cvar_i)^2} \ dt 
\end{align*}
for $i,j = 1, \ldots, p$.
The matrix $A$ is Hermitian while $B$ is symmetric (complex), so that
$\cnj B = B^\dagger$. This equation
can be reduced since it contains only half of the
unknowns as the initial residual level function. For example $\beta^\top
\cnj A\ \cnj \beta = \beta^\dagger A \beta$ and $\beta^\dagger B \cnj \beta
 = \overline{(\beta^\top \cnj B \beta)}$, so that
\begin{align*}
\begin{pmatrix}
\beta^\dagger & \beta^\top
\end{pmatrix}
\begin{pmatrix}
A & B \\ \cnj B & \cnj A 
\end{pmatrix}
\begin{pmatrix}
\beta \\ \cnj \beta
\end{pmatrix}
= 
2 \left[\beta^\dagger A \beta + \Re e \left(\beta^\dagger B \cnj \beta \right) \right]
\end{align*}
and similarly $\zeta^\dagger \beta = \overline{\zeta^\top \cnj \beta}$. With
the simplifications above the residual level function reduces to
\begin{equation}
\label{eq:reslev2}
F(\cvar,\beta) = \beta^\dagger A \beta + \Re e \left(\beta^\dagger B \cnj \beta
\right)  - 2 \Re e \left( \zeta^\dagger \beta\right) + {1 \over 2} \|h\|^2 .
\end{equation}

Now we can require the latter equation to satisfy the reflection
symmetry P. Equation~\eqref{eq:relab1} maps poles $\cvar_j$ and
coefficients $\beta_j$ respectively to $\CPPole_k$ and $\CPCoeff_k$, so
as to make the reflection symmetry explicit.
The residual level function now takes on a new block form
\begin{align}
\label{eq:blkreslev1}
F = 
\begin{pmatrix}
\CPCoeff^\dagger & (\pty \CPCoeff)^\top \end{pmatrix}
\begin{pmatrix}
X & Y \\ Y^\dagger & \pty{\cnj X} \end{pmatrix}
\begin{pmatrix}
\CPCoeff \\ \pty{\cnj \CPCoeff} \end{pmatrix}
& + \Re e \left[
\begin{pmatrix}
\CPCoeff^\dagger & (\pty \CPCoeff)^\top \end{pmatrix} 
\begin{pmatrix}
W & Z \\ Z^\dagger & \pty{\cnj W} \end{pmatrix} 
\begin{pmatrix}
\cnj \CPCoeff \\ \pty \CPCoeff \end{pmatrix} 
\right] \nonumber  \\  & - 2 \Re e \left[ 
\begin{pmatrix}
\theta^\dagger & (\pty \theta)^\top \end{pmatrix}
\begin{pmatrix}
\CPCoeff \\ \pty{ \cnj \CPCoeff} \end{pmatrix}
\right] + {1 \over 2} \|h\|^2
\end{align}
with
\begin{align*}
  W_{k,\ell} &= \int_{-\infty }^{+\infty} \frac{\WeightFct (t)}{(t- \cnj \CPPole_k)(t - \cnj \CPPole_\ell)} \ dt
  &
  \pty{\cnj W}_{k,\ell} &= \int_{-\infty }^{+\infty} \frac{\WeightFct (t)}{(t + \CPPole_k)(t + \CPPole_\ell)} \ dt
  \\
  X_{k,\ell} &= \int_{-\infty }^{+\infty} \frac{\WeightFct (t)}{(t - \cnj \CPPole_k)(t - \CPPole_\ell)} \ dt
  &
  \pty{\cnj X}_{k,\ell} &= \int_{-\infty }^{+\infty} \frac{\WeightFct (t)}{(t + \CPPole_k)(t + \cnj \CPPole_\ell)} \ dt
  \\
  Y_{k,\ell} &=  \int_{-\infty }^{+\infty} \frac{\WeightFct (t)}{(t- \cnj \CPPole_k)(t + \cnj \CPPole_\ell)} \ dt
  &
  Z_{k,\ell} &=  \int_{-\infty }^{+\infty} \frac{\WeightFct (t)}{(t- \cnj \CPPole_k)(t + \CPPole_\ell)} \ dt
  \\
  \theta_k &= \int_{-\infty }^{+\infty} \frac{\WeightFct (t) h(t) }{t- \cnj \CPPole_k} \ dt
  &
  \pty{\cnj \theta}_k &= \int_{-\infty }^{+\infty} \frac{\WeightFct (t) h(t) }{t + \CPPole_k} \ dt
\end{align*}
for $k,\ell = 1, \ldots, q$. These submatrix blocks preserve some of
the properties of the matrix they are part of. For instance,
$X^\dagger = X$ and $\cnj W = W^\dagger$ while the P symmetry imposes
new equalities $Y^\dagger = \pty{\cnj Y}$ and
$Z^\dagger = \pty{\cnj Z}$. In addition to these, the P symmetry
allows for additional equalities thanks to  the
symmetric integration boundaries of the inner product
$\langle \cdot, \cdot \rangle$. In other words
\begin{align*}
\int_0^a g(t,\CPPole_k)  = \int_{-a}^0 \pty{g}(t,\CPPole_k) 
& \qquad \textrm {for $g$ an even function of $t$} \\
\int_0^a g(t,\CPPole_k)  = - \int_{-a}^0 \pty{g}(t,\CPPole_k)
& \qquad \textrm {for $g$ an odd function of $t$} . 
\end{align*}
The direct implication of this observation is that, for instance,
$X=\pty{X}$, $W=\pty{W}$.

If we expand the matrix expression for the residual level function
and exploit all the symmetries the final
expression for $F$ becomes
\begin{align}
\label{eq:reslev3}
F = &\ \CPCoeff^\dagger  X  \CPCoeff +  \left(\CPCoeff^\dagger X \CPCoeff\right)^\top
- 2\ \Re e \left(\CPCoeff^\dagger Y
\cnj \CPCoeff\right) + \Re e \left[ \CPCoeff^\dagger W \cnj \CPCoeff +
\left(\CPCoeff^\dagger W \cnj \CPCoeff \right)^\dagger  -  2\ \CPCoeff^\dagger Z \CPCoeff \right] \\
\nonumber &- 2\ \Re e \left(
  \theta^\dagger \CPCoeff - (\pty \theta)^\top \cnj \CPCoeff \right) +
  {1 \over 2} \|h\|^2. 
\end{align} 
Despite the apparent complexity of the expression above, it would have
been quite more complex if we started to compute the residual level
function directly from Equation~\eqref{eq:ratfun2sym};  the quadratic
term in $\CPCoeff$s alone would have accounted for 16 terms. Moreover the
expression is a function of only $\CPCoeff_k$s and their conjugates.  $Z$
and $Y$ matrices appear only once, while $X$ and $W$ can be
transformed after their computation.


\subsection{Gradient of the Residual Level Function}
\label{sec:gradients}
Most optimization methods---including all the ones that we
consider---require the gradient of the residual level function.
We could compute the gradients analytically by using the expression
for $F$ derived in Equation~\eqref{eq:reslev1}. A simpler way is to
compute $\nabla f$ out of the formulation in
Equation~\eqref{eq:ratfunapp} in conjunction with the formulation of
$F$ in terms on \mbox{inner products $\langle \cdot, \cdot \rangle$}
\begin{equation}
\label{eq:gengrad}
\nabla F =\  \nabla \langle h - f , h - f \rangle = - \langle \nabla f
             , h - f \rangle - \langle h - f , \nabla f \rangle
=\ 2 \left[ \langle f , \nabla f \rangle - \langle h , \nabla f
    \rangle \right]. 
\end{equation}
Since $\cnj f = f$ and the $\nabla$ operator acts on the whole inner
products, the quantities $\langle \nabla f, f \rangle$ and
$\langle f , \nabla f \rangle$, while formally different, are actually
the same. In addition, the gradient of a real function with respect to
the conjugate of a complex variable is necessarily the conjugate of
the gradient with respect to variable itself: 
$\nabla_{\cnj z} f = {\partial f \over \partial \cnj z} = {\partial
  \cnj f \over \partial \cnj z} = \cnj{\nabla_{z}f}$.
Consequently we do not need to explicitly compute the gradient with
respect the conjugate of the poles and the
coefficients. Equation~\eqref{eq:gengrad} can be written making
explicit use of the C-symmetric or the full CP-symmetric formulation
of $f$, which is Equations~\eqref{eq:ratfun1sym} and
\eqref{eq:ratfun2sym} respectively.

\noindent
{\em C symmetry} -- Let us first derive $f$ with respect to
$\cvar_k$ and $\beta_k$
\[
{\partial f \over \partial \cvar_k} = {\beta_k \over (t - \cvar_k)^2}
\quad ; \quad 
{\partial f \over \partial \beta_k} = {1 \over (t - \cvar_k)}.
\]
The expression above is the $k$\textsuperscript{th}-component of the gradients
$\nabla_{\cvar}f$ and $\nabla_{\beta}f$ respectively.  Plugging them
in Equation~\eqref{eq:gengrad} and separating terms, we arrive
at the matrix expressions
\begin{align}
\nabla_{\cvar}F & =\ 2\left[ \beta^\dagger \nabla\!A + \beta^\top
  \nabla\!\cnj{B} - \nabla\!\zeta^\dagger \right] I_\beta \\
\nabla_{\beta}F & =\ 2\left[ \beta^\dagger A + \beta^\top
  \cnj{B} - \zeta^\dagger \right]
\end{align}
where $A$, $\cnj B$ and $\zeta$ are the same quantities defined in
Section~\ref{sec:3-1}, while remaining matrices are defined for $i,j = 1, \ldots, p$:
\begin{align*}
\nabla\!A_{i,j} &= \int_{-\infty }^{+\infty} \frac{\WeightFct (t)}{(t- \cnj \cvar_i)(t - \cvar_j)^2} \ dt  
&\qquad \nabla\!\zeta_i &= \int_{-\infty }^{+\infty} \frac{\WeightFct (t) h(t) }{(t- \cnj \cvar_i)^2} \ dt \\
\nabla\!\cnj{B}_{i,j} &=  \int_{-\infty }^{+\infty}
  \frac{\WeightFct (t)}{(t- \cvar_i)(t -  \cvar_j)^2} \ dt
& \quad I_\beta &= \diag (\beta) 
\end{align*}

\noindent
{\em CP symmetry} -- Analogously to what done in the C-symmetric
case, we first compute the derivatives of $f$ with respect to the poles
and the coefficients
\[
{\partial f \over \partial \CPPole_k} = \CPCoeff_k \left[
{1 \over (t - \CPPole_k)^2} + {1 \over (t + \CPPole_k)^2} \right]
\quad ; \quad 
{\partial f \over \partial \CPCoeff_k} = 
{1 \over (t - \CPPole_k)} - {1 \over (t + \CPPole_k)}.
\]
After entry-wise substitution of the above components of $\nabla_wF$
and $\nabla_\CPCoeff F$, some tedious rearrangement, and using the
parity with respect to the integration limits, we arrive at the
following matrix equations
\begin{align}
\nabla_wF & =\ 4 \left[ \CPCoeff^\dagger \left( \nabla\!X 
- \nabla\!Z 
\right) +
\CPCoeff^\top \left( \nabla\!\cnj{W} 
- \nabla\!\cnj{Y}
\right) - \nabla\!\theta^\dagger 
\right] I_\CPCoeff \label{eq:gradFomega}\\
\nabla_\CPCoeff F & =\ 4 \left[ \CPCoeff^\dagger \left( X  - Z -
\right) +
\CPCoeff^\top \left( \cnj{W} - \cnj{Y} -
\right) - \theta^\dagger \right]. \label{eq:gradFgamma}
\end{align}
The matrices not previously introduced are defined as follows.
For $k,\ell = 1, \ldots, q$:
\begin{align*}
\nabla\!X_{k,\ell}  &= \int_{-\infty }^{+\infty} \frac{\WeightFct(t)}{(t -
  \cnj \CPPole_k)(t - \CPPole_\ell)^2} \ dt &  \qquad\,
\nabla\!Z_{k,\ell}  &= - \int_{-\infty }^{+\infty}
\frac{\WeightFct (t)}{(t - \cnj \CPPole_k)(t + \CPPole_\ell)^2} \ dt \\
\nabla\!\cnj{W}_{k,\ell}  &= \int_{-\infty }^{+\infty} \frac{\WeightFct (t)}{(t - \CPPole_k)(t -
  \CPPole_\ell)^2} \ dt &  \qquad
\nabla\!\cnj{Y}_{k,\ell}  &=  - \int_{-\infty }^{+\infty}
  \frac{\WeightFct (t)}{(t - \CPPole_k)(t + \CPPole_\ell)^2} \ dt \\
\nabla\!\cnj{\theta}_{k}  &=  \int_{-\infty }^{+\infty}
  \frac{\WeightFct (t) h(t)}{(t - \CPPole_k)^2} \ dt
& \qquad I_\CPCoeff &= \diag (\CPCoeff) 
\end{align*}
Notice that, in both C-symmetric and CP-symmetric cases, the gradients
are row vectors. While this is an arbitrary choice, it is a quite
natural one to make. Moreover we decided to maintain the overall
multiplicative factor in front of them; such a factor was scaled out of
$F$ when writing this in a symmetry transparent form. Once again this
an arbitrary choice since re-scaling the gradient of $F$ does not
influence the minimization process.


%% file: numpole.tex
With the formulation of the residual level function and its gradient
from the previous section, we can now proceed to optimize the
\LSOP/ filters. Our goal is to minimize the residual level function
$F$. However, there are a number of possible approaches to the
minimization problem. Starting at a given position \(x^{(0)}\),
\emph{descent methods} minimize \(F(x)\)
by iteratively refining this position, such that
\(F(x^{(k+1)}) \leq F(x^{(k)})\).
Using the CP symmetries the minimization problem can be stated as
\begin{align} 
\label{eq:unconstrOptimization} 
\min_x \; F(x) = 
\min_x \;  || h - f(\CPPole, \CPCoeff) ||_2^2 ,
\end{align}
with the initial choice of parameters, or \emph{starting position}, of
the form
\(x^{(0)} = [\CPPole^{(0)} \; \CPCoeff^{(0)}] = [ \CPPole_1^{(0)},
..., \CPPole_q^{(0)}, \CPCoeff_1^{(0)}, ..., \CPCoeff_q^{(0)} ]\).

In Section~\ref{sec:graddescent} we present a first method to minimize
Equation~\eqref{eq:unconstrOptimization}.  Section~\ref{sec:lm}
introduces the Levenberg-Marquardt method, a more sophisticated solver
for Least-Squares problems.  \(F(x)\)
is non-convex in the poles \(\CPPole\)
of the filter function and thus a local minimum \(x^{*}\)
is not guaranteed to be a global minimum. Depending on the starting
position, descent methods, including the Levenberg-Marquardt method,
may produce results that are far from globally optimal.
Section~\ref{sec:choiceStartingPosition} deals with the choice of the
starting position in a way that mitigates this problem.  We address
the choice of weight function for the Least-Squares optimization in
Section~\ref{sec:weights}.  In some cases it is advantageous to
optimize more than just the squared differences between the filter and
the indicator function \(h(t)\);
Section~\ref{sec:constrOptimization} delves into constrained
optimization.  Section~\ref{sec:additional} concludes with a
discussion of a variety of \LSOP/ filters obtainable with the methods
we discuss throughout this section.

There are alternative approaches to the optimization problem.  A very
different method is Branch and Bound (B\&B) which, if successful,
finds the global minimum.  However, B\&B requires a convex
underestimation of the objective function instead of the gradient.  We
will not consider this method for two reasons. First,
its success highly depends on the ability
of finding a suitable underestimation, which is not an easy task.  Second, B\&B usually is
computationally more expensive than the methods we present. Since our
goal is to generate filters on the fly, we focus on cheaper methods.

\subsection{Optimization via Gradient Descent}
\label{sec:graddescent}
In this section we remark, through an illustrative example, on a
number of issues of the optimization process, which we use to
formulate some significant guidelines that are illustrated in the
following sections. The most basic descent method is \emph{gradient
  descent}. A gradient descent step drives down the residual level
function, at the current position, along a direction collinear with
its negative gradient. Given some mild assumptions, one eventually
arrives at a local minimum. For a given starting position \(x^{(0)}\)
the update step of the gradient descent method is
\begin{equation} 
\label{eq:desc} 
x^{(k+1)} = x^{(k)} + s\cdot \Delta
x^{(k)} = x^{(k)} - s\cdot \nabla_x F(x)\big|_{x=x^{(k)}} \;,\; s
\geq 0 .
\end{equation}
There are number of ways to select the \emph{step-length} \(s\)
in Equation~\eqref{eq:desc}.  While it is possible to set \(s\)
as constant, we can improve the convergence of the method by
approximating
\(s^{(k)} = \text{argmin}_{1 \geq u \geq 0} F(x^{(k)}+ u \cdot \Delta
x^{(k)})\)
at each iteration.  To this end, we use a backtracking line search
\cite{bv_cvxbook}: \(s\)
is initialized at each step with \(s=1\)
and then \(s\)
is halved until
\(F(x+\Delta x) < F(x) + \frac{s}{2} \nabla_x F(x)^\dagger \Delta x\).

\subsubsection{An Example}
\label{sec:graddescentexample}
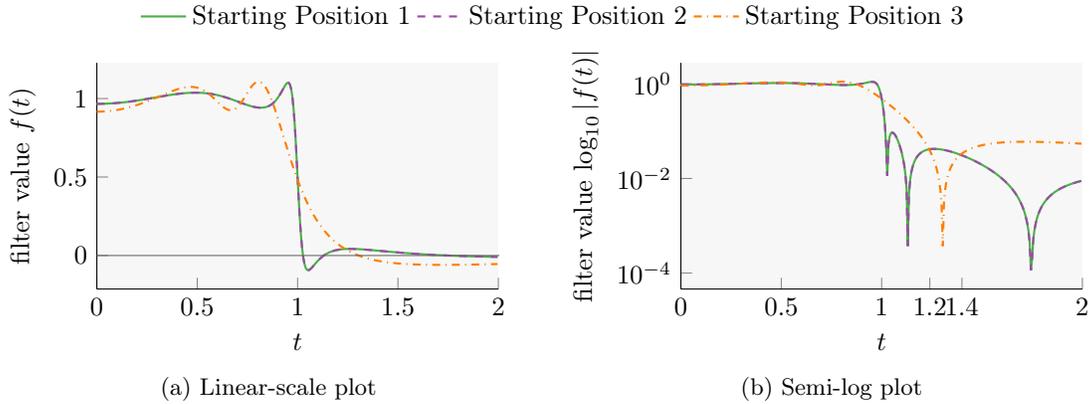
\begin{figure}
  \centering

  \ref{graddesc}


  \begin{subfigure}[b]{0.475\textwidth}
    \begin{tikzpicture}
      \pgfplotstableread{./FIGS/phi-SD.csv}{\fphiSD}
      \begin{axis}
        [ width=0.97\textwidth
        , height=0.2\textheight
        , axis y line*=left
        , axis x line*=bottom
        , restrict x to domain=0:2
        , xmin=0
        , xmax=2
        , axis background/.style={fill=black!3}
        , xlabel={$t$}
        , ylabel={filter value $f(t)$}
        , legend columns=-1
        , legend entries=
        { Starting Position 1
          , Starting Position 2
          , Starting Position 3
        }
        , legend to name=graddesc
        , legend style={draw=none}
        , cycle list name=exotic
        , ]
        \addplot+ [color=LSOPC,mark=none, thick] table [x index={0},y expr=\thisrowno{1}] {\fphiSD};
        \addplot+ [color=LSOPC2,dashed,mark=none, thick] table [x index={0},y expr=\thisrowno{2}] {\fphiSD};
        \addplot+ [color=LSOPC3,dashdotted,mark=none, thick] table [x index={0},y expr=\thisrowno{3}] {\fphiSD};
        \draw[color=XAXIS] (0,0) -- (5,0);
      \end{axis}
    \end{tikzpicture}
    \caption{Linear-scale plot}
  \end{subfigure}
  ~
  \begin{subfigure}[b]{0.475\textwidth}
    \begin{tikzpicture}
    \pgfplotstableread{./FIGS/phi-SD.csv}{\fphiSD}
      \begin{axis}
        [ width=0.97\textwidth
        , height=0.2\textheight
        , axis y line*=left
        , axis x line*=bottom
        , restrict x to domain=0:2
        , xmin=0
        , xmax=2
        , cycle list name=exotic
        , xtick= { 0, 0.5, 1, 1.24, 1.4, 2 }
        , x tick label style={/pgf/number format/fixed, /pgf/number format/precision=1}
        , ymode=log
        , axis background/.style={fill=black!3}
        , xlabel={$t$}
        , ylabel={filter value $\log_{10} |f(t)|$}
        , ]
        \addplot+ [color=LSOPC,mark=none, thick] table [x index={0},y expr=abs(\thisrowno{1})] {\fphiSD};
        \addplot+ [color=LSOPC2,dashed, mark=none, thick] table [x index={0},y expr=abs(\thisrowno{2})] {\fphiSD};
        \addplot+ [color=LSOPC3,dashdotted,mark=none, thick] table [x index={0},y expr=abs(\thisrowno{3})] {\fphiSD};
      \end{axis}
    \end{tikzpicture}
    \caption{Semi-log plot}
  \end{subfigure}
  \vspace*{-0.5cm}
  \caption{ Left side: Three \LSOP/ filters obtained from gradient descent with different starting positions.
    Right side: Semi-log plot with absolute value of the filter functions. The poles and residuals of the functions are listed in Table~\ref{tab:poles}.}
  \label{fig:sd_filter}
  \vspace*{-0.25cm}
\end{figure}

\begin{table}[t]
  \centering
  \begin{tabular}{llll}
    \toprule
    & Starting Position 1 & Starting Position 2 & Starting Position 3\\
    \midrule
    \(\CPPole_1\) & \(+0.997360 +0.044537 i\) & \(+0.721876 +0.550662 i\) & \(-0.742834 +0.246735 i\)\\
    \(\CPPole_2\) & \(+0.721876 +0.550662 i\) & \(+0.997360 +0.044537 i\) & \(+0.744336 +0.212182 i\)\\
    \midrule
    \(\CPCoeff_1\) & \(-0.024019 -0.002516 i\) & \(-0.135117 -0.148326 i\) & \(+0.791429 -0.331455 i\)\\
    \(\CPCoeff_2\) & \(-0.135117 -0.148326 i\) & \(-0.024019 -0.002516 i\) & \(+0.639334 +0.196414 i\)\\
    \midrule
    \(F([ \CPPole \; \CPCoeff])\) & \(0.005846\) & \(0.005846\) & \(0.033640\)\\
    \bottomrule
  \end{tabular}
  \caption{\label{tab:poles} Poles, coefficients, and residual levels
    of the filter functions show in Figure \ref{fig:sd_filter}}
\end{table}

In the following, we present three filters obtained using the
gradient descent method.  The real and imaginary part of the poles \(\CPPole\)
and coefficients \(\CPCoeff\) of the starting positions were chosen uniformly
random between \([-1,1]\)\footnote{The
  Least-Squares weights used to obtain these results are 1.0 inside
  \([-1000,1000]\)
  and 0.0 everywhere else.}.  For each filter function \(q=2\),
and so the filters have a total of 8 poles and coefficients. Using the
symmetries described in Section \ref{sec:2-1}, we need to optimize
only \(\CPPole_1, \CPPole_2, \CPCoeff_1\),
and \(\CPCoeff_2\).
The three filters are plotted on semi-log scale and on a normal scale
on the right hand side and the left hand side of
Figure~\ref{fig:sd_filter} respectively. Table \ref{tab:poles} shows
the poles, coefficients, and residual levels of the resulting filter
functions.  Starting Position 1 and 2 end in the same filter.
Starting Position 3 generates a different filter with a residual level
that is one order of magnitude higher than the others.  As the
semi-log plot shows, this last filter will likely perform
significantly worse than the first two. This can be understood by
noticing that the filters obtained from Starting Position 1 and 2 have
a smaller absolute value in \([1,1.21]\)
and---not fully visible in the plot---\([1.4,\infty]\).
From this simple example and many others we have run, we can comment
on a number of undesirable properties of the gradient descent method:

\begin{remark}[Slow convergence] 
\label{rem:convergence} 
Gradient descent is a fairly slow method, with linear
convergence behavior at best.  Even when equipped with a line-search
the number of iterations required to reach a minimum is often in the
millions.  We see in Section~\ref{sec:lm} how convergence can be
substantially improved by switching to a more advanced Least-Squares
optimization method.
\end{remark}

\begin{remark}[Dependence on starting position] 
\label{rem:startingPosition} 
The filter obtained from the optimization depends on the starting
position \(x^{(0)}\).
Different starting positions can result in very different filters with
different residual levels. Our experiments indicate that when
increasing \(q\)
the optimization finds more and more local minima, and their residual
levels compared to the best known solutions get worse.  Ideally, we
want a method allowing us to choose the starting position in a manner that
consistently yields good results.  In
Section~\ref{sec:choiceStartingPosition} we discuss an approach to
choosing the starting position which mitigates this problem.
\end{remark}

\begin{remark}[Lack of flexibility]   
\label{rem:constraint}
In order to obtain filters with distinct properties we need to modify
the optimization process accordingly. Section~\ref{sec:weights}
outlines a number of criteria for choosing the Least-Squares weight
function \(\WeightFct(t)\).
Furthermore, out-of-the-box gradient descent does not support
constrained optimization.  In Section~\ref{sec:constrOptimization}
we illustrate a number of ways to use optimization constraints on
filters.
\end{remark}

\subsection{Improving Convergence Speed}
\label{sec:lm}
We address Remark~\ref{rem:convergence} by introducing the
Levenberg-Marquardt method
\cite{marquardt1963algorithm,levenberg1944method} (LM).  This is a
solver for non-linear Least-Squares problems with a better convergence
rate than gradient descent. LM is an hybrid of the gradient
descent and the Gauss-Newton method (a good description of both
methods can be found in \cite{nnls}).  We present the Gauss-Newton
method first, then introduce the LM method. We give a formulation based on the inner product presented
in Equation~\eqref{eq:innpro}.

For the sake of clarity, we re-write the residual level function in
terms of
\(\xi([\alpha \; z], t) = h(t) - f(t,z,\alpha)\)
\[ F(x) = ||\xi(x)||_2^2 = \langle \xi(x), \xi(x)\rangle 
= \int_{-\infty}^{\infty} \WeightFct (t) |
  \xi(x,t) |^2 dt = \int_{-\infty}^{\infty} \WeightFct (t) |h(t) -
  f(t,z,\alpha) |^2 dt,
\]
where we indicate the collection of parameters $[z \; \alpha]$ with $x$.
The basis of the Gauss-Newton method is a linear approximation of \(\xi\)
\begin{equation}
  \xi(x+\Delta x) = \xi(x) +  \nabla \xi (x)\cdot \Delta x \doteq
  \xi + \sum_i {\partial \xi \over \partial x_i} \Delta x_i
\end{equation}
which, in the following, we refer to with the shortcut notation $\xi + \nabla \xi\cdot \Delta x$.
The Levenberg-Marquardt method aims at minimizing a linear
approximation of the residual level function by Gauss-Newton
iterates. By using the linear approximation of $\xi$, such a
requirement can be formulated as the minimization of the following
function with respect to $\Delta x$ 
\begin{align*}
  F(x+ \Delta x) & = ||\xi(x+ \Delta x)||_2^2 
        = \langle \xi, \xi \rangle + 2\langle \xi, 
        \nabla \xi\cdot\Delta x \rangle + \langle  
        \nabla \xi\cdot \Delta x, \nabla \xi\cdot \Delta x\rangle \\
                 & = F(x) + \nabla F(x)\cdot \Delta x + (\Delta x)^\dagger\cdot \langle 
                   \nabla\xi, \nabla\xi\rangle\cdot \Delta x . 
\end{align*}
Notice that the condition
$\cnj{\xi}(x+ \Delta x) = \xi(x+ \Delta x)$ implies that
$(\Delta x)^\dagger\cdot \cnj{\nabla}\xi = \nabla\xi\cdot\Delta x$
with the consequence that
$\langle \nabla\xi\cdot \Delta x, \xi\rangle = \langle\xi,
\nabla\xi\cdot\Delta x \rangle$. Taking the partial derivative of
$F(x+ \Delta x)$ with respect to $\Delta x$ and equating it to zero
one gets
\begin{equation*}
  \partial_{\Delta x} F(x+ \Delta x) = 2\left[\langle \xi, \nabla \xi  \rangle 
    + \langle \nabla \xi , \nabla \xi  \rangle\cdot\Delta x\right]  = 0 
\end{equation*}

Using the formulation above, the Gauss-Newton method iterates over
\(k\) in the following way:
\begin{align*}
  \text{1. Set: } & \quad H := 
                    \langle \nabla \xi(x^{(k)}) , \nabla \xi(x^{(k)})  \rangle \\
  \text{2. Solve:  } & \quad H \cdot \Delta x_{GN}^{(k)} 
                       = \langle \xi(x^{(k)}), \nabla\xi(x^{(k)}) \rangle  = - {1\over 2}\nabla F(x^{(k)}) \\
  \text{3. Update: }&\quad x^{(k+1)} = x^{(k)} + s \cdot \Delta
                      x_{GN}^{(k)} .
\end{align*}
Here $H$ acts as a linear approximation of the Hessian of $F$.  Consequently,
if $H$ is well-conditioned, the Gauss-Newton method can have quadratic
convergence.  However, there are still scenarios where Gauss-Newton
steps provide only a small improvement of the residual.  In those
cases it is beneficial to temporarily switch to gradient descent, even
though it only has linear convergence at best.  The
Levenberg-Marquardt method employs a \emph{dampening parameter}
\(\mu \geq 0\)
to switch between the Gauss-Newton and gradient descent methods.  By
adding a dampening term, the second step is substituted by
\begin{equation}
  \text{2. Solve: } \quad ( H + \mu I  ) \Delta x_{LM}^{(k)}  
  = - {1\over 2}\nabla F(x^{(k)}), 
\label{eq:LMSolve}
\end{equation}
where the gradient of $F$ is intended over both the poles and
coefficients.  Such an addition is equivalent to a constrained
Gauss-Newton with the parameter $\mu$ working as a Lagrange
multiplier.

In practice, the dampening parameter $\mu$ is re-adjusted after every
iteration.  For a small $\mu$, the solution to
Equation~\eqref{eq:LMSolve} is similar to the Gauss-Newton update step
$\Delta x_{LM} \simeq \Delta x_{GN}$. 
On the other hand, when $\mu$ is large then
$\Delta x_{LM} \simeq - \frac{1}{\mu} \nabla F(x)$, which is similar
to the gradient descent update step. Often, instead of $H + \mu I$
one uses $H + \mu \cdot \text{diag}(H)$.  Generally, when a
LM-step reduces the residual by a large amount, then $\mu$ is
decreased.  When a step does not decrease the residual, or does not
reduce the residual by enough, $\mu$ is increased.  So for
 large $\mu$ the resulting update $\Delta x_{LM}$ becomes
increasingly similar to the gradient descent update.

Since the formulations of the LM method given so far is independent
from the symmetry of $f$, it is valid for any formulation of the
rational filter. Let us look now more in detail at $H$. This is a
matrix whose entries are
$H_{i,j} = \langle \nabla_{x_i} f, \nabla_{x_j} f \rangle$ which, in
the case of CP symmetry, can be represented in block form as
\begin{equation}
  \label{eq:Hmatrix}
  H =
  \begin{pmatrix}
    H_2 & - \cnj{H}_1 & \cnj{H}_1 & -H_2\\
    -H_1 & H_2^\dagger & - H_2^\dagger & H_1 \\
    H_1 & - H_2^\dagger & H_2^\dagger & - H_1 \\
    - H_2 & \cnj{H}_1 & -\cnj{H}_1 & H_2
  \end{pmatrix}.
\end{equation}
Only two of the matrix blocks composing $H$ are independent and are
made of the following sub-blocks
\begin{equation}
  \label{eq:H1matrix}
  H_1 =
  \begin{pmatrix}
    \langle \nabla_\CPPole f, \nabla_\CPPole f \rangle & \langle \nabla_\CPPole f, \nabla_\CPCoeff f \rangle\\
    \langle \nabla_\CPCoeff f, \nabla_\CPPole f \rangle & \langle
    \nabla_\CPCoeff f, \nabla_\CPCoeff f \rangle
  \end{pmatrix}
  = 2
  \begin{pmatrix}
    I_\CPCoeff \left[ \nabla\nabla\!\cnj{W} - \nabla\nabla\!\cnj{Y}
    \right]
    I_\CPCoeff & I_\CPCoeff\left[ \nabla\!\cnj{W} - \nabla\!\cnj{Y} \right]^\top \\[2mm]
    \left[ \nabla\!\cnj{W} - \nabla\!\cnj{Y} \right] I_\CPCoeff &
    \cnj{W} + \cnj{Y}
  \end{pmatrix}
\end{equation}
\begin{equation}
  \label{eq:H2matrix}
  H_2 =
  \begin{pmatrix}
    \langle \nabla_{\cnj \CPPole} f, \nabla_\CPPole f \rangle &
    \langle \nabla_{\cnj \CPPole} f,
    \nabla_\CPCoeff f \rangle\\
    \langle \nabla_{\cnj \CPCoeff} f, \nabla_\CPPole f \rangle &
    \langle \nabla_{\cnj \CPCoeff} f, \nabla_\CPCoeff f \rangle
  \end{pmatrix}
  = 2
  \begin{pmatrix}
    I_\CPCoeff^\dagger \left[ \cnj{\nabla}\nabla\!X -
      \cnj{\nabla}\nabla\!Z \right]
    I_\CPCoeff & I_\CPCoeff^\dagger \left[ \nabla\!X - \nabla\!Z \right]^\dagger\\[2mm]
    \left[ \nabla\!X - \nabla\!Z \right] I_\CPCoeff & X - Z
  \end{pmatrix}
\end{equation}
where the only additional matrices that need to be defined are
\begin{align*}
  \nabla\nabla\!\cnj{W}_{i,j} = \int_{-\infty }^{+\infty} \frac{\WeightFct
  (t)}{(t- \CPPole_i)^2(t - \CPPole_j)^2} \ dt & \qquad
                                                 \nabla\nabla\!\cnj{Y}_{i,j} = - \int_{-\infty }^{+\infty} \frac{\WeightFct
                                                 (t)}{(t- \CPPole_i)^2(t + \CPPole_j)^2} \ dt \\
  \cnj{\nabla}\nabla\!X_{i,j} = \int_{-\infty }^{+\infty} \frac{\WeightFct (t)}{(t - \cnj{\CPPole}_i)^2(t- \CPPole_j)^2} \ dt & \qquad
                                                                                                                                \cnj{\nabla}\nabla\!Z_{i,j} = - \int_{-\infty }^{+\infty} \frac{\WeightFct (t)}{(t - \cnj{\CPPole}_i)^2(t+ \CPPole_j)^2} \ dt . 
\end{align*}
The factor of 2 in front of the block matrices $H_1$ and $H_2$ comes
from the P symmetry in combination with the evenness of the integral
boundaries (i.e. $X=\pty X$, $\nabla\!X = \nabla\!\pty{X}$, etc.).
Notice that block rows 3 and 4 of Eq.~\eqref{eq:Hmatrix} are exactly
equivalent, up to a sign, to block rows 2 and 1
respectively. Similarly, block columns 3 and 4 are proportional to
block columns 2 and 1. This is expected due to the CP symmetry. In
addition, $H_1$ is complex symmetric ($H_1^\dagger = \cnj{H}_1$), and
$H_2$ is complex Hermitian ($H_2^\dagger = H_2$) which, implicitly,
verifies that $H$ is complex Hermitian itself.

Going back to the Equation~\eqref{eq:LMSolve}, one can also write the
vector $\Delta x_{LM}$ in block notation,
\[
  (\Delta x_{LM})^\top = [(\Delta y)^\top\, - (\Delta y)^\dagger\,
  (\Delta y)^\dagger \, - (\Delta y)^\top]
\]
with $(\Delta y)^\top = [\Delta \CPPole^\top\, \Delta \CPCoeff^\top]$.
Writing Equation~\eqref{eq:LMSolve} in terms of $\Delta y$ with
$\mu \cdot \text{diag}(H)$ in place of $\mu I$ one obtains 4 separate
equations. Thanks to the symmetries of $H$, it is possible to reduce
them to a single equation solving for $\Delta y$. There are several
different way to achieve this result. Starting from the third
equation, one can extract $\left(H_1 \Delta y\right)$ as a function of $H_2^\dagger$,
$\Delta \cnj{y}$, and $\nabla_{\cnj y}F$, then take the conjugate of
it, and substitute it in the fourth equation. The final result is a
linear system solves whose dimension is one fourth of the original
size of $H$ 
\begin{equation} 
\label{eq:hessian}
\left[i\, \Im m\left(H_2\right) - \mu\, \text{diag}\left(H_2\right)\right]
\Delta y = {1\over 2} \nabla_y F
\end{equation}
The Levenberg-Marquardt method is faster than gradient descent.
An improved
convergence rate, together with the reduced size of the equation to
be solved, substantially accelerate the time to solution. This is a
critical step for being able to generate filters on the fly.

\subsection{Systematic Choice of Starting Position}
\label{sec:choiceStartingPosition}
In this section we address the dependence of the optimization process
on the initial conditions.  In Remark~\ref{rem:startingPosition} we
pointed out that the optimized filter crucially depends on the
starting position.  Usually, the choice of starting positions for
non-linear optimization is a difficult problem.  Many heuristics
exist; some methods rely on solving a convex variant of the problem
\cite{bv_cvxbook}.  Generally, it is preferable to use domain
knowledge in the initial choice of starting position. Recall that the
parameters of our optimization, the poles and coefficients, define a
rational filter function.  In general, random choices for \(\CPPole\)
and \(\CPCoeff\)
do not produce good filters. We can reinterpret the problem of finding
good starting positions for the poles and coefficients as finding a
filter that serves as a good initial guess for an optimized filter.
To this purpose we could use contour filters expressed through a
numerical implementation of Cauchy's Residue Theorem (see
Sec.\ref{sec:1-2}), namely the Gauss and Trapezoidal filters,
or more specialized ones, like the Elliptic filter, that have already
been proposed as substitutes for the contour
solvers~\cite{2014arXiv1407.8078G}.

To realize why the starting positions are crucial for the optimization
method let us consider what happens when increasing $q$, the number of
poles and coefficients in one quadrant of $\CC$. For some spectrum
slices of a given eigenvalue problem, larger values for $q$, say
$q=10$ or higher, may be required to obtain the desired convergence of
the solver. For instance, some methods, such as the DD-PP projection
method \cite{KalantzisDD-PP} that do not iterate the subspace, have
little choice but to choose large $q$. A given value of $q$ results in
$4q$ real degrees of freedom in the optimization: $q$ poles and $q$
coefficients, each one with a real and imaginary part. As a result,
optimization with large $q$ becomes more and more expensive. In
Figure~\ref{fig:sd_filter} we have seen an example of multiple local
minima of the residual level function $F$ with $q=2$.  Our extensive
numerical experiments indicate that higher $q$ values not only
increase the number of local minima in $F$, but have the
effect of worsening the residual levels of most of these additional local minima w.r.t the best known local minimum.
In one of our numerical experiments, using a random starting position with $q=8$ yields a filter with a residual
of $5.4 \times 10^{-4}$, while with $q=4$ we achieved a residual level
of $2.4 \times 10^{-4}$.  Choosing $q=8$  requires 18 linear system
solves for each filter application, whereas \(q=4\) requires only 8 solves.
To offset the additional cost of the linear system solves we want a filter that has a lower residual level, thus likely requiring fewer subspace iterations.
Without a good starting position we obtain a filter with \(q=8\) that requires \emph{more} linear system solves and \emph{more} iteration.
What is more, the optimization for a filter with \(q=8\) is significantly more expensive that for \(q=4\).
Simply put, random starting positions yield worse and worse filters for larger \(q\).

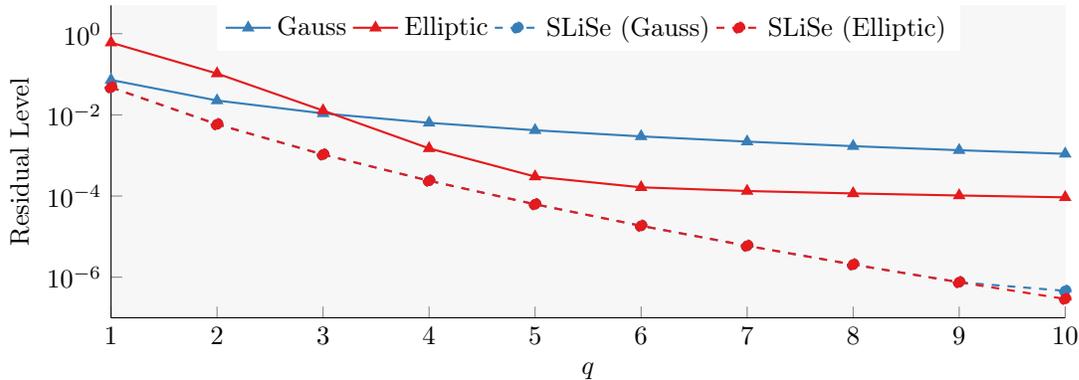
\begin{figure}
  \centering
  \begin{tikzpicture}
    \pgfplotstableread{data/resids.csv}{\residsfile}
    \begin{axis}
      [ axis background/.style={fill=black!3}
      , xlabel=$q$
      , ylabel=Residual Level
      , cycle list name=exotic
      , ymode=log
      , ymin=1e-7
      , ymax=5
      , width=.95\textwidth
      , height=0.25\textheight
      , xtick={1, ..., 10}
      , ytick={1,1e-2,1e-4,1e-6,1e-8}
      , xmin=1
      , xmax=10
      , legend style=
      { draw={none}
        , at={(current bounding box.north-|current axis.north)}
        , anchor=north
        , legend columns=-1
      }
      , axis y line*=left
      , axis x line*=bottom
      , ]
      \addplot [thick, color=GAUSSC, mark=triangle*] table [x=N,y={Gauss}] {\residsfile};
      \addlegendentry{Gauss}

      \addplot [thick, color=ZOLOC, mark=triangle*] table [x=N,y={Zolo}] {\residsfile};
      \addlegendentry{Elliptic}

      \addplot [thick, color=GAUSSC, mark=*, dashed] table [x=N,y={LS1}] {\residsfile};
      \addlegendentry{\LSOP/ (Gauss)}

      \addplot [thick, color=ZOLOC, mark=*, dashed] table [x=N,y={LS2}] {\residsfile};
      \addlegendentry{\LSOP/ (Elliptic) }
    \end{axis}
  \end{tikzpicture}
  \vspace*{-0.25cm}
  \caption{Residual levels of Gauss and Elliptic filters, as well
    as, \LSOP/ filters obtained from Gauss and Elliptic filters as
    starting positions.}
  \label{fig:startingPosGaussZolo}
  \vspace*{-0.25cm}
\end{figure}

While choosing random starting positions does not work well with
larger $q$, choosing existing filters produces very good results.
Figure \ref{fig:startingPosGaussZolo} shows the residual
levels\footnote{The weight function used has unit weight between -1000
  and 1000} of different filters for different $q$.  The residual
levels of the Gauss filter are indicated with blue triangles, the
residual levels of the Elliptic filter in red triangles.  Shown with
blue circles is the residual level of the filter that is obtained when
the initial position of the \LSOP/ optimization is started with the
Gauss filter poles and coefficients.  Conversely, the red circles
indicate the residual levels achieved by the filter obtained by
starting \LSOP/ optimization with the Elliptic filter.
The \LSOP/ filters have significantly lower residual levels than the
Gauss and Elliptic filters. Using poles and coefficients from Gauss
and Elliptic filters as starting positions for a $q < 10$ results
in the same unconstrained \LSOP/ filter.  Although the Gauss and
Elliptic are quite different, such result indicates that both of them
are in the convex region around the same minimum.  The number of
iterations in the LM method differs for both starting positions, when
starting with the Elliptic filter fewer iterations are required.
While we cannot claim optimality for these filters, at least in the
unconstrained case, we have never obtained a filter besting the filters
obtained by choosing the poles and coefficients of an Elliptic filter
as starting positions.
%
In our experience existing filters, and the Elliptic filter in
particular, make for excellent starting positions practically
eliminating the problem of finding an appropriate starting position.

\subsection{\texorpdfstring{The Least-Squares Weights \(\WeightFct(t)\)}{The Least-Squares Weights}}
\label{sec:weights}

\newcolumntype{F}{>{\centering\arraybackslash}p{1.8cm}}%
\newcolumntype{E}{>{\centering\arraybackslash}p{1.7cm}}%
\newcolumntype{G}{>{\centering\arraybackslash}p{0.7cm}}%
\newcolumntype{H}{>{\centering\arraybackslash}p{1.4cm}}%
\newcolumntype{I}{>{\centering\arraybackslash}p{1.0cm}}%

\begin{table}[t]
\begin{center}
\small
\begin{tabularx}{\textwidth}{ X  I  E  F  E  H  I  G  I}
\toprule
$|t|\in$  & $[0, .95)$  & $[.95, .995)$ & $[.995, 1.005)$ 
  & $[1.005, 1.05)$ & $[1.05, 1.2)$ & $[1.2, 2)$ 
  & $[2, 5)$ & $[5, \infty)$ \\
\midrule
  $\WeightFct_{1}(t)$ & 1 & 4 & 2 & 4 & 1 & 0 & 0 & 0  \\
  $\WeightFct_{2}(t)$ & 1 & 4 & 2 & 4 & 1 & .05 & .001 & 0 \\
  $\WeightFct_{3}(t)$ & 1 & 4 & 0 & 0 & 1 & .2 & .2 & 0 \\
\bottomrule
\end{tabularx}
  \caption{Three separate weight functions \(\WeightFct_1, \WeightFct_2\), and \(\WeightFct_3\). The resulting filter functions have undesirable properties that make them poor rational filters.}
  \label{fig:wtFct}
\end{center}
\end{table}

The attentive reader might have noticed that we have not yet discussed
what causes \(\gamma\)-\LSOP/
and \(\eta\)-\LSOP/
from Section \ref{sec:results} to be different filters.  Both filters
were obtained using the LM optimization method and the Elliptic filter
as a starting position.  A major difference between the two is
the Least-Squares weight function $\WeightFct(t)$---``weights'', for
short---used in the optimization process.  As discussed in
Remark~\ref{rem:constraint} the optimization process requires some
flexibility.  The weight function supplies this flexibility, as
confirmed by the very different properties of \(\gamma\)-\LSOP/
and \(\eta\)-\LSOP/.

The result of the optimization is invariant under scaling of the
weight function so, for the sake of consistency, we consider scaled
weight functions with $\WeightFct(0) = 1$.
With the
flexibility provided by the weight function comes the problem of
specifying a weight function that results in a filter with the desired
properties. We will see that setting weights without a specific strategy
almost certainly yields unexpected results.

We aim for a fire-and-forget approach for the optimization process.
In this sense, our target is to obtain the desired results without
parameter space exploration. This section contains three guidelines
for the choice of the weight functions. The guidelines 
 help the user with the choice of the weights outside the search
interval $[-1,1]$, near its endpoints, and inside it.
Each guideline is a heuristic that can be easily implemented as
part of an optimization algorithm, or checked manually.  These guidelines are an important
ingredient for a usable optimization, as following them
virtually eliminates the problem of filters with unexpected properties.
We provide examples for unexpected properties that occur when violating Guideline 1 and 2.
However, if
a filter with very specific properties---such as the \(\gamma\)-\LSOP/
or \(\eta\)-\LSOP/
filters---is desired, some trial-and-error is unavoidable.

The weight function must be non-zero in an interval larger than just
\([-1,1]\).  Otherwise the optimization may result in filters that
have large absolute values outside $[-1, 1]$.  However, care must be
taken when choosing the weights outside the search interval.  By
tapering the weights too quickly we obtain a steep filter that will
nevertheless exhibit insufficient convergence rates.  In this context
the quality of a filter can be expressed as a function of the filters
local extrema outside of the search interval.  Specifically,
non-increasing local maxima of the absolute function value away from
the search interval is a very desirable property.  Existing filters
already have this property. The Gauss filter, for example, has quickly
decaying local extrema.  
\begin{guideline}[Outside]
  Slowly taper off the weights in a large enough neighborhood outside
  the search interval. In order to avoid trial-and-error, we suggest
  to detect increasing local maxima of the absolute value of the rational function during
  the optimization process and increase the weights where required.
\end{guideline}

By construction, existing filters have a value of 0.5 at the endpoints of the search
interval, $f(1) = f(-1) = 0.5$.  While not strictly
required for most contour solvers, other methods that use rational
filters do require a value of 0.5 at the end of the search interval.
One example of such a method is the estimation of eigenvalues inside
an interval \cite{1308.4275}.  \LSOP/ filters are not guaranteed to
have a value of 0.5 at the endpoints, unless an appropriate
constrained is added to the optimization.
In practice \LSOP/ filters with \(f(\pm 1) \approx 0.5\)
are easily achievable via appropriate weight functions. Such an
approximate value is enough for most applications. By choosing the weight
function as symmetric around some region around the endpoint of the search interval
we achieve \(f(\pm 1) \approx 0.5\).
If an exact value is required the filter can be scaled accordingly.
\begin{guideline}[Endpoints]
  The weight function around the endpoints of the search interval
  should be chosen symmetrically. Accordingly, during the
  optimization process, choose
  $\WeightFct(\pm 1-\epsilon) = \WeightFct(\pm 1+\epsilon)$ for
  $\epsilon \leq 0.2$.  Additionally, the optimization procedure should check the
  weight function for symmetry around the search interval endpoints
  and, if desired, scale the resulting filter to $f(\pm 1) = 0.5$.
\end{guideline}

When the weights inside $[-1, 1]$ are too small the filter function
will oscillate inside the search interval.  Such oscillations cause
the filter value to dip below 1.0, hurting the convergence rate of the
entire subspace. We can effectively solve this problem by adjusting
the weight function during the optimization process.  Previously, we
suggested to increase the weights outside the search interval to
produce filters with non-increasing local maxima.  By the same token
we can detect any oscillation inside $[-1, 1]$ and increase the
assigned weights accordingly.
\begin{guideline}[Inside]
  The weights in the entire search interval should be chosen large
  enough to prevent large oscillations. During the optimization
  process, monitor the difference between minima and maxima of the
  filter inside $[-1, 1]$ and adapt weights accordingly.
\end{guideline}

\begin{figure}[t]
  \centering

  \ref{wtlsop}

  \begin{subfigure}[b]{0.475\textwidth}
    \begin{tikzpicture}
      \pgfplotstableread{./data/wt.csv}{\fphiSD}
      \begin{axis}
        [ width=0.97\textwidth
        , height=0.2\textheight
        , axis y line*=left
        , axis x line*=bottom
        , restrict x to domain=0:3
        , xmin=0
        , xmax=3
        , axis background/.style={fill=black!3}
        , xlabel={$t$}
        , ylabel={filter value $f(t)$}
        , legend columns=-1
        , legend entries=
        { \(\WeightFct_1\)
          , \(\WeightFct_2\)
          , \(\WeightFct_3\)
        }
        , legend to name=wtlsop
        , legend style={draw=none}
        , cycle list name=exotic
        ]
        \addplot+ [color=LSOPC,mark=none, thick] table [x index={0},y expr=\thisrowno{1}] {\fphiSD};
        \addplot+ [color=LSOPC2,dashed,mark=none, thick] table [x index={0},y expr=\thisrowno{2}] {\fphiSD};
        \addplot+ [color=LSOPC3,dashdotted,mark=none, thick] table [x index={0},y expr=\thisrowno{3}] {\fphiSD};
        \draw[color=XAXIS] (0,0) -- (5,0);

      \end{axis}

      \begin{axis}
        [ width=0.55\textwidth
        , height=0.12\textheight
        , axis y line*=left
        , axis x line*=bottom
        , xmin=0.95
        , xmax=1.1
        , ymin = -0.1
        , xtick={0.95,1,1.025,1.1}
        , minor xtick={1.05}
        , axis background/.style={fill=white}
        , every tick label/.append style={font=\tiny}
        , forget plot
        , xshift=2.5cm
        , yshift=1.50cm
        ]
        \addplot+ [color=LSOPC,mark=none, thick] table [x index={0},y expr=\thisrowno{1}] {\fphiSD};
        \addplot+ [color=LSOPC2,dashed,mark=none, thick] table [x index={0},y expr=\thisrowno{2}] {\fphiSD};
        \addplot+ [color=LSOPC3,dashdotted,mark=none, thick] table [x index={0},y expr=\thisrowno{3}] {\fphiSD};
        \draw[color=XAXIS] (0,0) -- (5,0);

        \coordinate (o) at (0,-0.1);
        \coordinate (inter1) at (1.025,0.5);
        \draw[dotted, color=LSOPC3!80] (inter1|-o) -- (inter1) --  (inter1-|o);

      \end{axis}
    \end{tikzpicture}
    \caption{Linear-scale plot. \\Smaller plot: Abscissa limited to \([0.9,1.1]\).}
    \label{fig:wt:linlin}
  \end{subfigure}
  ~
  \begin{subfigure}[b]{0.475\textwidth}
    \begin{tikzpicture}
      \pgfplotstableread{./data/wt.csv}{\fphiSD}
      \begin{axis}
        [ width=0.97\textwidth
        , height=0.2\textheight
        , axis y line*=left
        , axis x line*=bottom
        , restrict x to domain=0:3
        , xmin=0
        , xmax=3
        , ymin=1e-4
        , cycle list name=exotic
        , xtick= { 0, 1, 1.3, 2, 2.5, 3 }
        , x tick label style={/pgf/number format/fixed, /pgf/number format/precision=1}
        , ymode=log
        , axis background/.style={fill=black!3}
        , xlabel={$t$}
        , ylabel={filter value $\log_{10} |f(t)|$}
        , ]
        \addplot+ [color=LSOPC,mark=none, thick, name path=lsop1graph] table [x index={0},y expr=abs(\thisrowno{1})] {\fphiSD};
        \addplot+ [color=LSOPC2,dashed, mark=none, thick] table [x index={0},y expr=abs(\thisrowno{2})] {\fphiSD};

        \coordinate (o) at (0,1e-4);
        \coordinate (inter1) at (1.3,0.03);
        \draw[dotted, color=LSOPC2!80] (inter1|-o) -- (inter1) --  (inter1-|o);

        \coordinate (inter2) at (2.5,0.03486);
        \draw[dashdotted, color=LSOPC2!90] (inter2|-o) -- (inter2) -- (inter2-|o);

      \end{axis}
    \end{tikzpicture}
    \caption{Semi-log plot. \\\(\WeightFct_3\) omitted for better readability.}
    \label{fig:wt:loglin}
  \end{subfigure}

  \vspace*{-0.5cm}
  \caption{Plots of filters obtained with three separate weight functions \(\WeightFct_1, \WeightFct_2\), and \(\WeightFct_3\). The resulting functions act as poor rational filters.}
  \label{fig:wt}
  \vspace*{-0.25cm}
\end{figure}
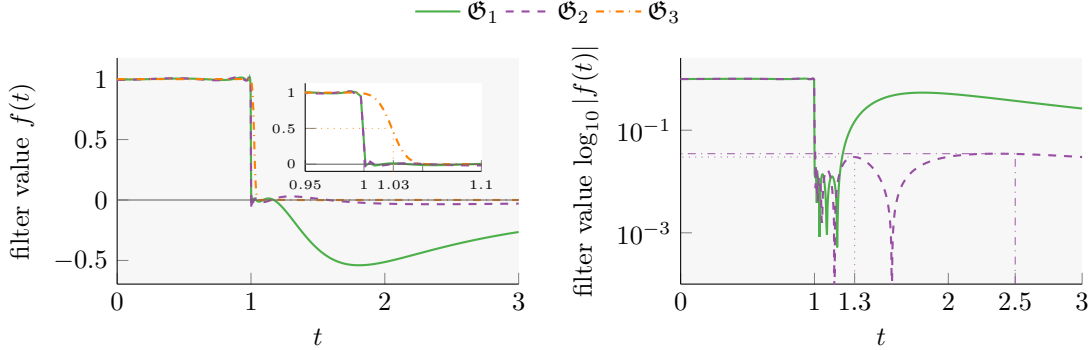

\paragraph{Example 1: Very Small Weights Outside the Search Interval}
Let us look what happens when the resulting filters were
obtained without adhering to one of the three guidelines.
In the process of obtaining a replacement candidate for the Elliptical filter the weight function in Table~\ref{fig:wtFct} could be considered.
We will see that $\WeightFct_{1}$ and $\WeightFct_{2}$  do not correctly taper off the weights outside
$[-1, 1]$ and thus violate Guideline 1.

Starting from the assumption that only the weights near the endpoints
are relevant, so $\WeightFct_{1}$ is chosen with non-zero values only
for $|t| < 1.2$.  Figure \ref{fig:wt} shows the filter function
obtained with $\WeightFct_{1}$.  The filter is a good approximation
within $[-1.2,1.2]$. However, for $|t| > 1.2$, the rational function
poorly approximates zero. Such a function would hardly work as a
filter. In fact, a value of $f(1.8) \approx -0.5$ for the filter would
result in severely degraded convergence rates (see
Equation~\eqref{eq:convratio}).  To obtain a better filter it is
enough to adjust the weight function to have non-zero values in a
larger interval around $[-1,1]$, for instance $[-5,5]$. The weight
$\WeightFct_2$ satisfy such a requirement, but the weights taper off
very quickly for $t >|1.2|$.  The resulting filter function (see
Figure \ref{fig:wt}) has values close to zero outside the search
interval. Despite this improved behavior, the filter has another more
subtle shortcoming. Figure \ref{fig:wt:loglin} shows in a semi-log
plot that the filter drops steeply to low values near $[1,1.2]$, just
as desired. However, the filters values inside $[1.2,3]$ are
\emph{larger} than the values inside $[1,1.2]$. %
In other words, some of the local maxima outside of the search
interval increase away from it, resulting once again in deteriorated
convergence rates. Large enough weights in the interval
$[1.2 < |t| < 5]$ will result in non-increasing local minima.


\paragraph{Example: Asymmetric Weights Near Endpoints}

The weight $\WeightFct_3$ in Table~\ref{fig:wtFct} illustrates the
results of a weight function that is not symmetric around the endpoint of the
search interval, violating the second guideline. In particular,
$\WeightFct_3$ has a larger weight on the inside of the search interval than
on the outside. Figure~\ref{fig:wt:linlin} plots the resulting filter
function as an inlay inside the figure. 
Assigning a weight of zero between $0.995$ and $1.05$ results in a
filter that fails to drop to zero at the end of the search interval: its
value at the endpoint is almost one!  Because of the delayed drop the
filter crosses the value $0.5$ at a later point
$f_{\WeightFct_{3}}(1.025) \approx 0.5$. This behavior does not appear
in the filter produced with $\WeightFct_1$ and $\WeightFct_2$ as
they respect the second guideline: the weights are symmetric around
the endpoints $|t| = 1$ and $f(\pm 1) \approx 0.5 $ without the need
to re-scale the filter. This effect can be understood in terms of
the minimization of the squared errors during the function
approximation.  The continuous drop of the filter function at the
endpoints, no matter how steep, incurs in a large squared error.  When
the weight function lacks symmetry around the endpoint of the search
interval, the optimization shifts the drop towards the part of the
neighborhood of $[-1, 1]$ with lesser weight.  The result is a filter
value $f( \pm 1) < 0.5 $ if the weight is smaller just inside the
search interval, and a value $f(\pm 1) > 0.5 $ if the weight is smaller
just outside the search interval.  Choosing symmetric
weights around the endpoints results in $f(\pm 1) \approx 0.5$

\subsection{Constrained Optimization}
\label{sec:constrOptimization}

\begin{figure}
  \centering
  \ref{plotpenlsop}

  \begin{subfigure}{.475\textwidth}
    \centering
    \begin{tikzpicture}
      \pgfplotstableread{./data/p6-2e-10.csv}{\fphiSD}
      \begin{axis}
        [ width=.97\textwidth
        , height=.2\textheight
        , xmin=0.9999
        , xmax=1.0001
        , axis y line*=left
        , axis x line*=bottom
        , xlabel={$t$}
        , ylabel={fitler value $f(t)$}
        , axis background/.style={fill=black!3}
        , xtick = {0.99995, 1, 1.00005}
        , x tick label style={/pgf/number format/fixed , /pgf/number format/precision=6}
        , legend columns=-1
        , legend entries=
        { \LSOP/ \texttt{c=+0.000}
          , \LSOP/ \texttt{c=-pen}
          , \LSOP/ \texttt{c=+pen}
        }
        , legend to name=plotpenlsop
        , legend style={draw=none}
        ]

        \addplot+ [color=LSOPC,mark=none, thick] table [x index={0},y expr=\thisrowno{2}] {\fphiSD};

        \addplot+ [color=LSOPC2,mark=none, dashed, thick] table [x index={0},y expr=\thisrowno{3}] {\fphiSD};

        \addplot+ [color=LSOPC3,mark=none,dashdotted, thick] table [x index={0},y expr=\thisrowno{4}] {\fphiSD};

        \draw[color=XAXIS] (0.995,0) -- (1.005,0);
      \end{axis}
    \end{tikzpicture}
    \caption{\texttt{pen}$=2\cdot10^{-10}$ }
    \label{fig:p6-plusminus}
  \end{subfigure}
  \centering
  ~
  \begin{subfigure}{.475\textwidth}
    \begin{tikzpicture}
      \pgfplotstableread{./data/p6-2e-9.csv}{\fphiSD}
      \begin{axis}
        [ width=0.97\textwidth
        , height=0.2\textheight
        , xmin=0.999
        , xmax=1.001
        , ymin=-0.2
        , ymax=1.2
        , axis y line*=left
        , axis x line*=bottom
        , xlabel={$t$}
        , ylabel={filter value $f(t)$}
        , axis background/.style={fill=black!3}
        , legend style={at={(0.5,-0.1)}, anchor=north}
        , scaled x ticks = false
        , x tick label style={/pgf/number format/fixed, /pgf/number format/precision=4}
        , forget plot
        ]

        \addplot+ [color=LSOPC,mark=none, thick] table [x index={0},y expr=\thisrowno{2}] {\fphiSD};

        \addplot+ [color=LSOPC2,mark=none,dashed, thick] table [x index={0},y expr=\thisrowno{3}] {\fphiSD};

        \addplot+ [color=LSOPC3,mark=none,dashdotted, thick] table [x index={0},y expr=\thisrowno{4}] {\fphiSD};

        \draw[color=XAXIS] (0.995,0) -- (1.005,0);

      \end{axis}

      \begin{axis}
        [ width=0.45\textwidth
        , height=0.12\textheight
        , xmin=0.999
        , xmax=1.001
        , ymin=-5
        , ymax=5
        , axis y line*=left
        , axis x line*=bottom
        , axis background/.style={fill=white}
        , every tick label/.append style={font=\tiny}
        , scaled x ticks = false
        , xtick = {0.999, 1, 1.001}
        , x tick label style={/pgf/number format/fixed, /pgf/number format/precision=4}
        , forget plot
        , xshift=0.5cm
        , yshift=0.75cm
        ]

        \addplot+ [color=LSOPC,mark=none, thick] table [x index={0},y expr=\thisrowno{2}] {\fphiSD};

        \addplot+ [color=LSOPC2,mark=none,dashed, thick] table [x index={0},y expr=\thisrowno{3}] {\fphiSD};

        \addplot+ [color=LSOPC3,mark=none,dashdotted, thick] table [x index={0},y expr=\thisrowno{4}] {\fphiSD};

        \draw[color=XAXIS] (0.995,0) -- (1.005,0);
      \end{axis}

    \end{tikzpicture}
    \caption{\texttt{pen}$=2\cdot10^{-9}$ }
    \label{fig:p6-bad2}
  \end{subfigure}
  \vspace*{-0.5cm}
  \caption{Examples of filters obtained via penalty terms. Left: an appropriately sized penalty parameter yields filters of different steepness. Right: Too large a (positive) penalty parameter results in a pole on the real axis.}
  \label{fig:pen}
  \vspace*{-0.25cm}
\end{figure}
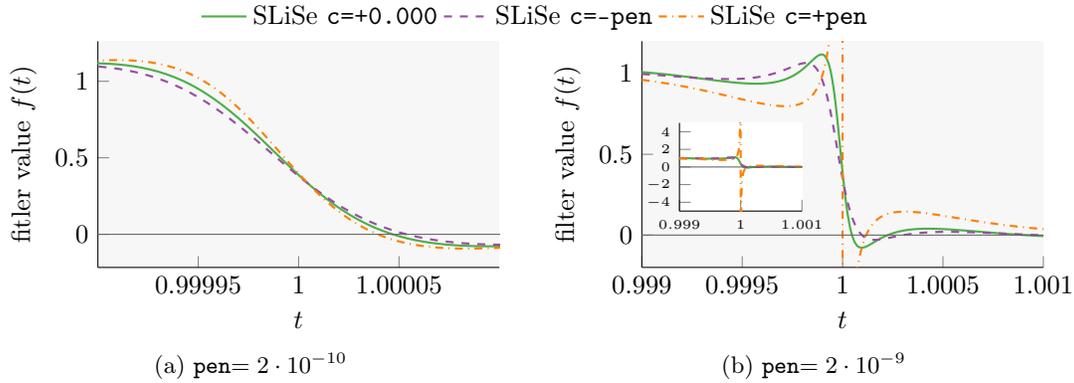

We have seen that the weight function provides some flexibility to the
optimization process. However, some desirable properties of a filter
cannot be directly influenced by it. In this section
we discuss two filter properties that can be achieved by implementing
constraints on the optimization. First, we discuss a penalty term on
the gradient of the filter function, which results in filters of
varying 'steepness' and with varying amounts of 'overshooting'. Given
our standard search interval, \([-1,1]\),
we call the derivative of the filter \(f(t)\)
at the endpoints of the interval
\mbox{\(\nabla_t f(t) |_{t=1} = - \nabla_t f(t) |_{t=-1}\)}
the \emph{steepness} of the filter.  The steepness---or separation
factor---of a filter has previously been suggested as a quality
measure in the context of optimizing exclusively the coefficients of
the rational function~\cite{saad16LeastSquares}.  A steep filter is
considered desirable because it dampens eigenvalues outside the search
interval even when the eigenvalues are close to the interval endpoints
and, in doing so, improves convergence.  When optimizing the poles of
the rational functions together with its coefficients the steepness
alone does not serve as a good measure of quality anymore.
It is possible for a filter to be steep, and yet not to dampen
eigenvalues outside of the search interval sufficiently well.  The
lack of dampening is due to an effect called ``overshooting'',
corresponding to a filter that greatly exceeds its target value.  Overshooting
is a known behavior when approximating discontinuous functions with
continuous ones.  
Usually a very steep filter tends to overshoot at the endpoint of
$[-1 1]$, and conversely filters that do not overshoot are not
particularly steep.

In Section \ref{ref:penalty} we discuss how to adjust the optimization
to generate filters of varying steepness, and thus also with varying
amounts of overshooting.  A \emph{penalty term} can modify the
optimization procedure to influence the steepness.  By adding a term
to the residual level function we penalize filters that are not
desirable by a factor of \(c\), the penalty parameter.
\begin{align}
  \label{eq:penalty}
  \min_{\CPPole_i, \CPCoeff_i \; 1 \leq i \leq q }& || h - f(\CPPole,
  \CPCoeff) ||_2^2 + {c\over 2} \cdot \left[\nabla_t f(t)|_{t=1} 
  - \nabla_t f(t)|_{t=-1}\right] .
\end{align}
Note that since $\nabla_t f(1)$ is supposed to be negative, a positive
$c$ will make for a steeper filter when minimizing the residual.
Solving the optimization problem requires straight-forward changes to
the gradient.


Section \ref{sec:boxconstraints} describes a different kind of
constraint, namely box constraints on the imaginary part of the poles.
Such a constraint may be advantageous when Krylov solvers are used for
the linear system solves as they appear in Equation~\eqref{eq:linsolve}.
The condition number of the shifted matrix $A-\CPPole_i I$ may become
large when $|\Im m(\CPPole_i)|$ is small, which can occur for large
degrees of the Elliptic and \LSOP/ filters.  One possible way to
circumvent this problem is to ensure that the absolute value of the
imaginary part for each pole is large enough.  For \LSOP/ filters this
translates into a box constraint \(d\) on the parameters:

\begin{equation}
  \label{eq:boxconstr}
  \min_{\CPPole_i, \CPCoeff_i \; 1 \leq i \leq q } || h(t) - f(t, \CPPole, \CPCoeff) ||_2^2
  \quad \textrm{s.t.}\quad \forall i:  |\Im m\{ \CPPole_i \}| \geq d \ 
\end{equation}


Both constrained optimization approaches presented in this section are
not fire-and-forget; some trial and error is always required for the
parameter selection.  This lack of robustness for the parameter
selection is the exception in our work. 
Constrained optimization without the need for extensive
parameter selection is an area of future work.

\subsubsection{Steepness and Overshooting via Penalty Term}\label{ref:penalty}

\begin{table}[t]
  \centering
  \begin{tabular}{c c c}
    \toprule
    \(c\) & Residual & Steepness\\
    \midrule
    0 & \(2.265829 \times 10^{-5}\) & \(-1.211081 \times 10^{4}\)\\
    \(-2 \times 10^{-10}\) & \(2.279018 \times 10^{-5}\) & \(-1.063877 \times 10^{4}\)\\
    \(+2 \times 10^{-10}\) & \(2.288353 \times 10^{-5}\) & \(-1.429542 \times 10^{4}\)\\
    \bottomrule
  \end{tabular}
  \caption{\label{tab:cpm} Residual levels and steepness of the filter functions shown in Figure \ref{fig:p6-plusminus}.}
\end{table}

Before we look at some examples of filters obtained with penalty
terms, let us consider how the steepness and overshooting influence
the convergence behavior of the filter by comparing \(\gamma\)-\LSOP/
and the Gauss filter from Figure \ref{fig:sd_filter:nonlog}. By
starting to drop significantly earlier than the endpoint of the search
interval at $t=1$, the Gauss filter is significantly less steep than
the \LSOP/ filter. Conversely, the \LSOP/ filter maintains a value
close to one inside the search interval closer to the endpoint, and it
takes small values just outside the search interval.  The increased
steepness of the $\gamma$-\LSOP/ filter comes at the cost of
overshooting.  Figure \ref{fig:sd_filter:nonlog} clearly shows that
the \LSOP/ filter shoot up above 1 and drops below 0 just inside and
outside the search interval respectively.
In terms of the convergence rate, the surge above 1 is not a problem,
however the overshooting outside $[-1 1]$ is an
undesirable effect. An overshooting filter maps eigenvalues to
relatively large negative---and thus large in magnitude---values
outside the endpoints, hurting convergence.

While it is possible to increase the steepness of \LSOP/ filters via
the penalty term, there is little benefit in even steeper filters with
more overshooting.  Figure~\ref{fig:p6-plusminus} shows an
illustrative example of filters obtained with an appropriately sized
penalty parameter and number of poles per quadrant $q=4$. Table
\ref{tab:cpm} shows the residuals without penalty term and the
steepness of the filter $\nabla_t f(1)$ for the resulting filters.
The figure makes plain how a negative penalty term tends to decrease
steepness and overshooting while a positive $c$ increases them. Since
the penalty parameter is chosen significantly smaller than the residual, the
change to the filter is not very pronounced.  Choosing the penalty
parameter to be larger is not without hazards, especially when $c$ is
chosen to be positive. Figure~\ref{fig:p6-bad2} shows a filter
generated with a larger value for $c$, again with $q=4$. Such a filter
has a pole on the real axis which results in a very steep slope, but
not an effective filter. For one thing, the pole causes significant
overshooting which substantially hurts convergence. Additionally,
a real pole affects the corresponding linear system, which can become
seriously ill-conditioned causing failure or slow converge of the
iterative solver.  While a large and positive parameter is problematic
because of the risk of a real pole, this is usually not a concern in
practical applications. Often, it is more important to limit the
overshooting than it is to make the already steep \LSOP/ filters even
steeper.  Figure~\ref{fig:p6-bad2} shows that a negative penalty
parameter $c$ reduces overshooting\footnote{Some trial and error may
  be required to obtain a penalty parameter of appropriate size.}.
\(\eta\)-\LSOP/ was obtained using a negative penalty parameter, which
we discuss further in Section \ref{sec:additional}.

\subsubsection{Large Imaginary Parts via Box Constraints}\label{sec:boxconstraints}
Implementing box constraints in our framework is
straightforward: Instead of the old update step in Equation
\eqref{eq:desc} we use gradient projection by updating as follows
\begin{equation} \label{eq:gradientProjection} x^{(k+1)} = \mathcal P(
  x^{(k)} - s\cdot \nabla_x F(x)\big|_{x=x^{(k)}} ) \;,\; s \geq 0 ,
\end{equation}
where $\mathcal P(x)$ projects $x$ into the constraints. In practice
we project onto the constrained value whenever the constraint is
violated.  We implement this scheme by forcing the absolute value of
the imaginary part of every pole to be equal or larger than some value
\texttt{lb}. For a single pole \(\CPPole_l\) the projection would be
\begin{equation*}
  \mathcal P(\CPPole_l)=
  \begin{cases}
    \Re e(\CPPole_l) + i\cdot \text{sgn}(\Im m(\CPPole_l)) \cdot \text{\texttt{lb}} & |\Im m(\CPPole_l)| < \text{\texttt{lb}}\\
    \Re e(\CPPole_l) + i\cdot \Im m(\CPPole_l) &\text{ otherwise }
  \end{cases}
\end{equation*}

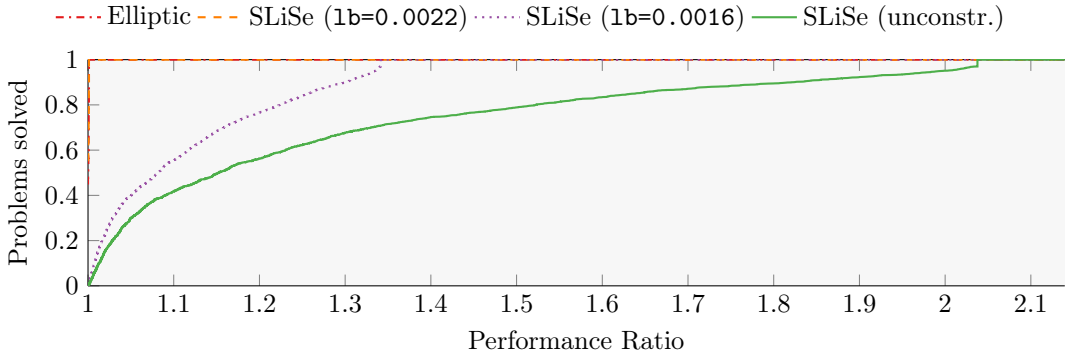
\begin{figure}
  \centering
  \ref{lsoplb}
\begin{tikzpicture}
  \begin{axis}[
    enlargelimits=false,
    axis y line*=left,
    ymin=0, ymax=1,
    ytick={0,0.2,0.4,0.6,0.8,1.0},
    xlabel={Performance Ratio}, ylabel={Problems solved},
    width=.97\textwidth,
    height=0.2\textheight,
    axis background/.style={fill=black!3},
    , legend columns=-1
    , legend entries=
    { Elliptic
      ,\LSOP/ (\texttt{lb=0.0022})
      ,\LSOP/ (\texttt{lb=0.0016})
      ,\LSOP/ (unconstr.)
    }
    , legend to name=lsoplb
    , legend style={draw=none}
    ]

  \addplot+[mark=none, thick, color=ZOLOC, dashdotted] table [x=x, y=y] {data/box.tex_Elliptic};
  \addplot+[mark=none, thick, color=LSOPC3, dashed] table [x=x, y=y] {data/box.tex_imagellip};
  \addplot+[mark=none, thick, color=LSOPC2, dotted] table [x=x, y=y] {data/box.tex_imaghalf};
  \addplot+[mark=none, thick, color=LSOPC] table [x=x, y=y] {data/box.tex_unconstr};



  \end{axis}
\end{tikzpicture}
\caption{Performance profile comparing the worst condition number for the linear systems arising from filter application to 2116 problems obtained from the ``Si2'' matrix.
  The performance ratio for the Elliptic and \LSOP/ (\texttt{lb=0.0022}) filters are one over the entire abscissa range.}
    \label{fig:lbPP}
  \vspace*{-0.25cm}
\end{figure}

In the following, we compare filters with different box constraints.
Figure \ref{fig:lbPP} shows four filters each with \(q=4\).
Shown are the Elliptic filter, a \LSOP/ filter obtained via
unconstrained optimization\footnote{The weights are available in
  Appendix~\ref{sec:app:filterlist} under ``Box-\LSOP/''.}, and two
additional \LSOP/ filters obtained with different \texttt{lb}
constraints.  For the Elliptic filter
\(\min_{1\leq i \leq q} |\Im m (\CPPole_i)|\)
is about \(0.0022\),
while it is \(0.001\)
for the unconstrained \LSOP/ filter.  In the case of the constrained
\LSOP/ filters, we consider ``\LSOP/ (\texttt{lb=0.0022})'' filter, with a
\texttt{lb} corresponding the the Elliptic filter.  Additionally, we
also examine ``\LSOP/ (\texttt{lb=0.0016})'' generated with a
constraint of 0.0016, about half a way between the Elliptic and the
unconstrained filter.  We calculate the condition number for each of
the shifted linear systems that result from filter application on the
2116 problems extracted from the ``Si2'' matrix as illustrated in
Section~\ref{sec:results}.  Figure \ref{fig:lbPP} is a performance
profile of the largest condition number for each of the benchmark
problems.  The Elliptic filter and the \LSOP/ (\texttt{lb=0.0022})
perform identically in the performance profile.  Both filters have a
value of one over the entire abscissa range, which indicates that the
condition numbers are the same.  The figure implies that the largest
condition number is influenced solely by the smallest absolute
imaginary value of the poles.  Accordingly, the constrained filter with
\texttt{lb} of 0.0016 performs worse, and the unconstrained \LSOP/
filter is the worst.  This effect can be understood by realizing
that the shifted matrices $A - I\CPPole_l$ become nearly singular only
if $\CPPole_l$ is near an eigenvalue of $A$.

An increase in the \texttt{lb} constraint has a large influence on the filter.
As a result, caution is required when optimizing with box constraints.
For example, for a choice of weight function that follow Guideline 1
from Section~\ref{sec:weights} in the unconstrained case, the same is
not necessarily true when optimizing a filter with box constraints.
We discuss an example filter with box constraints in the next section.

\subsection{A Rich Variety of Filters: Practices and Experience}
\label{sec:additional}

I the previous sections we have discussed a number of techniques to influence the
optimization procedure.  Now we are going to illustrate three filters
that highlight the potential of these optimization parameters.  First,
we discuss  \(\eta\)-\LSOP/,
where a penalty parameter is used to limit overshooting.  Second, we
present a filter that violates Guideline 1 from
Section~\ref{sec:weights} and yet exhibits good rates of convergence
on our set of benchmark problems.  Third, we present a filter that
uses box constraints to achieve better condition numbers for the
shifted matrices that arise from the filter application. Despite the
Gauss filter is the one most often used in real-world applications, in this section
we put a strong emphasis on alternatives for the Elliptic filter.
The Elliptic filter is an
interesting case study, because its rate of convergence is worst-case
optimal.  We focus on improving over the Elliptic filter to
illustrates the power of \LSOP/ filters.

\paragraph{Penalty parameter: \(\eta\)-\LSOP/}

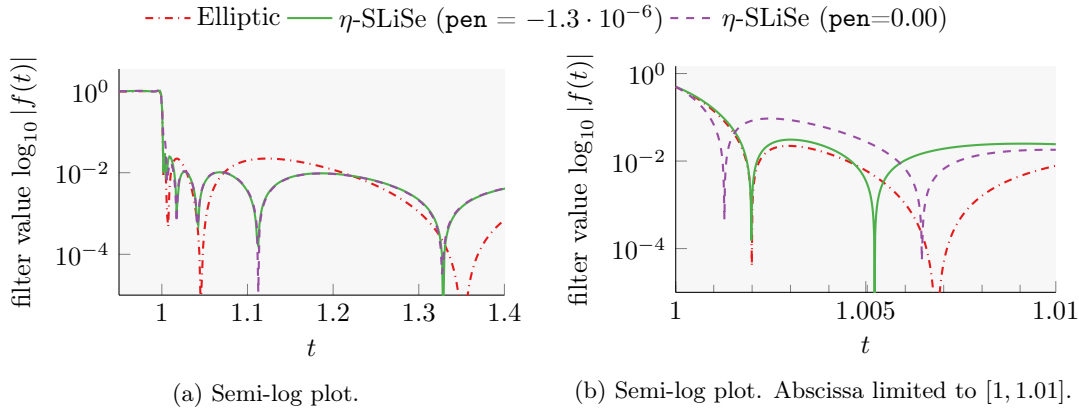
\begin{figure}
  \centering
  \ref{plotetalog}

  \begin{subfigure}[b]{.47\textwidth}
    \centering
    \begin{tikzpicture}
      \pgfplotstableread{./data/eta2.csv}{\fphiSD}
      \begin{axis}
        [ width=0.95\textwidth
        , height=0.2\textheight
        , axis y line*=left
        , axis x line*=bottom
        , restrict x to domain=0:2
        , xmin=0.95
        , xmax=1.4
        , ymin=1e-5
        , cycle list name=exotic
        , ymode=log
        , axis background/.style={fill=black!3}
        , xlabel={$t$}
        , ylabel={filter value $\log_{10} |f(t)|$}
        , ]
        \addplot+ [color=ZOLOC,dashdotted,mark=none, thick] table [x index={0},y expr=abs(\thisrowno{3})] {\fphiSD};
        \addplot+ [color=LSOPC,mark=none, thick] table [x index={0},y expr=abs(\thisrowno{2})] {\fphiSD};
        \addplot+ [color=LSOPC2,dashed, mark=none, thick] table [x index={0},y expr=abs(\thisrowno{1})] {\fphiSD};
      \end{axis}
    \end{tikzpicture}
    \caption{Semi-log plot. \\}
    \label{fig:etacomp:Nozoom}
  \end{subfigure}
  ~
  \begin{subfigure}[b]{.47\textwidth}
    \centering
    \begin{tikzpicture}
      \pgfplotstableread{./data/eta.csv}{\fphiSD}
      \begin{axis}
        [ width=0.94\textwidth
        , height=0.2\textheight
        , axis y line*=left
        , axis x line*=bottom
        , restrict x to domain=0:2
        , xmin=1
        , xmax=1.01
        , ymin=1e-5
        , cycle list name=exotic
        , xtick= { 1, 1.005, 1.01 }
        , minor xtick={1, 1.001, ..., 1.01}
        , x tick label style={/pgf/number format/fixed, /pgf/number format/precision=3}
        , ymode=log
        , axis background/.style={fill=black!3}
        , xlabel={$t$}
        , ylabel={filter value $\log_{10} |f(t)|$}
        , legend columns=-1
        , legend entries={Elliptic,\(\eta\)-\LSOP/ (\texttt{pen} = \(-1.3 \cdot 10^{-6}\)),$\eta$-\LSOP/ (\texttt{pen}=0.00)}
        , legend to name=plotetalog
        , legend style={draw=none}
        , ]
        \addplot+ [color=ZOLOC,dashdotted,mark=none, thick] table [x index={0},y expr=abs(\thisrowno{3})] {\fphiSD};
        \addplot+ [color=LSOPC,mark=none, thick] table [x index={0},y expr=abs(\thisrowno{2})] {\fphiSD};
        \addplot+ [color=LSOPC2,dashed, mark=none, thick] table [x index={0},y expr=abs(\thisrowno{1})] {\fphiSD};
      \end{axis}
    \end{tikzpicture}
    \caption{Semi-log plot. Abscissa limited to \([1, 1.01]\).}
    \label{fig:etacomp:Zoom}
  \end{subfigure}
  \vspace*{-0.5cm}
  \caption{Plots of the Elliptic filter, \(\eta\)-\LSOP/, and the
    \(\eta\)-\LSOP/ filter without penalty term.  All filters have 16
    poles and coefficients.}
  \label{fig:etacomp}
  \vspace*{-0.25cm}
\end{figure}

We have seen \(\eta\)-\LSOP/
in Section~\ref{sec:results:eta}, where we compare its convergence
rate to the Elliptic and Gauss filters.  \(\eta\)-\LSOP/
was obtained via a negative penalty parameter, to limit the
overshooting of the filter.  Without the penalty term the filter
would perform significantly worse.

Figure~\ref{fig:etacomp} shows the Elliptic and \(\eta\)-\LSOP/
filter.  Additionally, we include the unconstrained version of
\(\eta\)-\LSOP/,
being the filter obtained with the same weights as \(\eta\)-\LSOP/
but without penalty term.  Figure~\ref{fig:etacomp:Nozoom} shows
an abscissa range of \([0.95,1.4]\).
At this scale both \LSOP/ filters look identical.  Both filters
oscillate less than the Elliptic filter, which we
would expect from the rate of convergence for \(\eta\)-\LSOP/.
Figure~\ref{fig:etacomp:Zoom} shows the same filters for an abscissa
range of \([1,1.01]\),
just outside the search interval.  At this scale, the two \LSOP/
filters are very different.  The \LSOP/ filter without penalty
term is steeper and overshoots more than the Elliptic filter,
which results in larger absolute function values inside
\([1.001,1.006]\).
The penalty parameter for \(\eta\)-\LSOP/
was chosen large enough to make the filter about as steep as the
Elliptic filter.  As a result, \(\eta\)-\LSOP/
overshoots less than the unconstrained filter and oscillates at a
magnitude only slightly larger than the Elliptic filter.  
Compared to the Elliptic filter, \(\eta\)-\LSOP/
trades-off a slightly larger (absolute) function values near the end
of the search interval for much smaller values farther away.

In this example, the penalty parameter is chosen to be of large
absolute value, as compared to the values in
Section~\ref{ref:penalty}.  Such a large penalty parameter results in
a significant reduction in the steepness and overshooting of the
filter.  The difference between the \LSOP/ filters with and without
penalty term is large near the end of the search interval, but
negligible in \([1.05, \infty]\).
The ``design procedure'' for such a filter is straight forward:
first, we chose the weights such that the filter has the desired
behavior for most of the \(t\)
axis, e.g., \([0,1] \cup [1.05, \infty]\)
and such that the filter overshoots slightly more than desired.  Then,
we used a negative penalty term to lessen the overshooting; this step
usually requires only very few iterations and so can be done very
quickly.

\paragraph{Breaking Guideline 1: \(\zeta\)-\LSOP/}

We specifically advocate 3 guidelines for the choice of weight
function in Section~\ref{sec:weights}.  However, when done judiciously
breaking these guidelines can be advantageous, as we illustrate in
this section.  We discuss \(\zeta\)-\LSOP/
a filter that violates Guideline 1; this filter also aims to replace
the Elliptic filter.

\begin{figure}
  \centering
  \ref{plotzetalsop}

  \begin{subfigure}[b]{.48\textwidth}
    \centering
    \begin{tikzpicture}
      \begin{axis}
        [ width=.98\textwidth
        , height=0.2\textheight
        , enlargelimits=false
        , ymin=0
        , ymax=1
        , xmin=1
        , xmax=20
        , axis background/.style={fill=black!3}
        , ytick={0.1,0.3,0.5,0.7,0.9}
        , xtick={1,10, 20}
        , minor xtick={1, 2, ..., 10, 11, 12,13,14,15,16,17,18,19}
        , extra x tick style={/pgfplots/major tick length=.2cm,/pgfplots/tick style={line width=1.5pt}}
        , xlabel={Performance Ratio}
        , ylabel={Problems solved}
        , axis y line*=left
        , axis x line*=bottom
        , y tick label style={/pgf/number format/.cd, fixed, precision=2, /tikz/.cd  }
        , legend columns=-1
        , legend entries={Gauss, Elliptic, \(\zeta\)-\LSOP/, \(\kappa\)-\LSOP/}
        , legend to name=plotzetalsop
        , legend style={draw=none}
        , xmode=log
        , log ticks with fixed point
        ]

        \addplot+[mark=none, thick, color=GAUSSC, dashed, name path=gaussgraph] table [x=x, y=y]   {data/perfprof_Si2.mat_04_zeta.tex_Gauss};
        \addplot+[mark=none, thick, color=ZOLOC, dashdotted, name path=zolograph] table [x=x, y=y] {data/perfprof_Si2.mat_04_zeta.tex_Elliptic};
        \addplot+[mark=none, thick, color=LSOPC, name path=lsopgraph] table [x=x, y=y]             {data/perfprof_Si2.mat_04_zeta.tex_Zeta};
        \addplot+[mark=none, thick, color=LSOPC2, name path=zolograph, draw=none] coordinates {(1,1)};

        \coordinate (o) at (1,0);

      \end{axis}
    \end{tikzpicture}
    \caption{The same performance profile zoomed into \([1,20]\)}
    \label{fig:zololike_4:zoom}
  \end{subfigure}
  ~
    \begin{subfigure}[b]{.47\textwidth}
    \begin{tikzpicture}
      \begin{axis}
        [ width=.97\textwidth
        , height=0.2\textheight
        , enlargelimits=false
        , restrict x to domain=0:2
        , xmin=1, xmax=2
        , ymin=0, ymax=1
        , axis background/.style={fill=black!3}
        , ytick={0.1,0.3,0.5,0.7, 0.9}
        , xtick={1,1.5,2}
        , minor xtick={1,1.1,...,2}
        , xlabel={Performance Ratio}
        , ylabel={Problems solved}
        , axis y line*=left
        , axis x line*=bottom
        , y tick label style={ /pgf/number format/.cd, fixed, precision=2, /tikz/.cd }
        , extra x tick style={/pgfplots/major tick length=.2cm,/pgfplots/tick style={line width=1.5pt}}
        ]

        \addplot+[mark=none, thick, color=GAUSSC, dashed, name path=gaussgraph] table [x=x, y=y] {data/perfprof_Si2.mat_04_kappa.tex_Gauss};
        \addplot+[mark=none, thick, color=ZOLOC, dashdotted] table [x=x, y=y] {data/perfprof_Si2.mat_04_kappa.tex_Elliptic};
        \addplot+[mark=none, thick, color=LSOPC2] table [x=x, y=y] {data/perfprof_Si2.mat_04_kappa.tex_Kappa};

        \coordinate (o) at (1,0);



      \end{axis}
    \end{tikzpicture}
    \caption{The same performance profile zoomed into the interval $[1,2]$.}
    \label{fig:perfprof_Si2_04_wc_krylov_zoom}
  \end{subfigure}
  \vspace*{-0.5cm}
  \caption{Semi-log performance profile of the convergence rate for
    the Gauss, $\zeta$-\LSOP/, and Elliptic filters. All filters have
    16 poles and $\mathfrak p = \mathfrak m$. Both filters violate Guideline 1.}
  \label{fig:zololike_4}
  \vspace*{-0.25cm}
\end{figure}
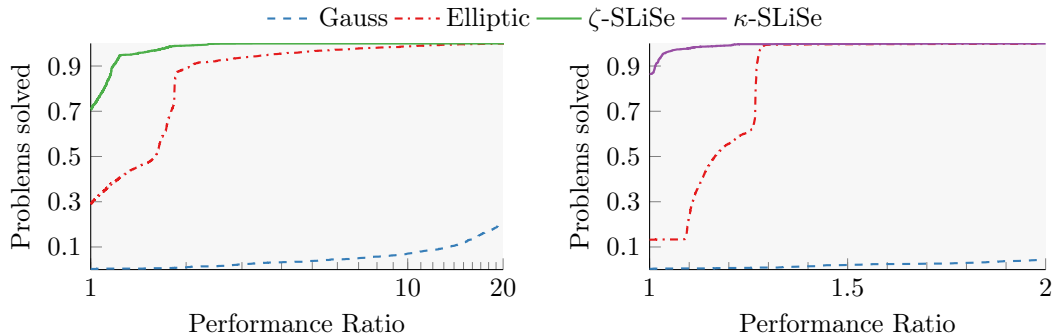

Figure~\ref{fig:zololike_4:zoom} shows a performance profile of the
Gauss, Elliptic, and \(\zeta\)-\LSOP/
filter that compares the rates of convergence for these filters.  The
setup is the same as in Section~\ref{sec:results:eta}: we use the same
2116 intervals taken from the ``Si2'' matrix,
\(\mathfrak p = \mathfrak m\),
and all filters have 16 poles. The Gauss filter performs similar to
the performance profile in Section~\ref{sec:weights}.
\(\zeta\)-\LSOP/
filter not only performs better than the Elliptic filter, it appears
to perform better than \(\eta\)-\LSOP/!
\(\eta\)-\LSOP/
achieves a better rate of convergence than the Elliptic filter for
only 50\% of the problems, whereas \(\zeta\)-\LSOP/
does so for 70\% of the problems.  The performance profile also
encodes a more subtle difference: In Figure~\ref{fig:zololike_4:zoom}
\(f_{Elliptic}(2.35) \approx 0.92\),
which indicates that \(\zeta\)-\LSOP/
exhibits rates of convergence better than 2.35 times the rate of the
Elliptic filter for 8\% of the benchmark problems.  This is not the
case for \(\eta\)-\LSOP/, where \(f_{Elliptic}(2.35) \approx 0.999\).

It appears that \(\zeta\)-\LSOP/
would be a better replacement for the Elliptic filter.  However, the
filter does not conform to Guideline 1.  The filter has a single
local extremum that has a significantly larger absolute value than
both of its neighboring extrema.  This extremum is still smaller than
the equi-oscillation of the Elliptic filter, which is the reason the
why \(\zeta\)-\LSOP/
performs better than the Elliptic filter.  The 8\% of benchmark
problems where \(\zeta\)-\LSOP/
performs significantly better than the Elliptic filter are exactly
those were no eigenvalue coincides with this larger local extremum of
\(\zeta\)-\LSOP/.
It is possible to obtain filters that violate our guidelines and still
yield good convergence rates, as illustrated by \(\zeta\)-\LSOP/.
However, adhering to our guidelines vastly simplifies the creation of
filters.  For that reason, we propose \(\eta\)-\LSOP/,
and not \(\zeta\)-\LSOP/, as a replacement for the Elliptic filter.


\paragraph{Box Constraint: \(\kappa\)-\LSOP/}

There is another problem with \(\zeta\)-\LSOP/.
The filter has a pole with an absolute imaginary value of
\mbox{\(\min_{1\leq i \leq q} |\Im m (\CPPole_i)| \approx 0.001\)}.
Analogously to the filters discussed in
Section~\ref{sec:boxconstraints} this deteriorates the convergence of
Krylov-based linear system solvers.  To mitigate this problem we can
use box constraints to obtain a filter where each pole has an absolute
imaginary value of at least 0.0022, the corresponding value for the
Elliptic filter.  We discuss a filter called \(\kappa\)-\LSOP/,
obtained with a box constraint of \texttt{lb}\(=0.0022\)
and the same weight function used to obtain \(\zeta\)-\LSOP/.
\(\kappa\)-\LSOP/ violates Guideline 1, just as \(\zeta\)-\LSOP/ does.

Figure~\ref{fig:perfprof_Si2_04_wc_krylov_zoom} shows the performance
profile for \(\kappa\)-\LSOP/.
The box constraint results in two changes of \(\kappa\)-\LSOP/
as compared to the \(\zeta\)-\LSOP/.
First, the box constraint contribute to lessen the overshooting of the
filter.  As a result \(\kappa\)-\LSOP/
attains a better rate of convergence than the Elliptic filter for 86\%
of the benchmark problems, as compared to the 70\% of
\(\zeta\)-\LSOP/.
Second, \(\kappa\)-\LSOP/
has larger (as compared to \(\zeta\)-\LSOP/)
absolute function values for most of the \(t\)
axis, which results in rates of convergence that are closer to the
rates of the Elliptic filter.  The larger absolute function values
cause Figure~\ref{fig:perfprof_Si2_04_wc_krylov_zoom} to look roughly
like a version of Figure~\ref{fig:zololike_4:zoom} compressed along
the abscissa.  From the performance profile we can infer that the
convergence rates of \(\kappa\)-\LSOP/
are only slightly better than those of the Elliptic filter.

Section~\ref{sec:boxconstraints} already illustrates that box
constraints can effectively decrease the condition number of the
linear systems that arise in the application of the filter.  The
condition numbers of the linear systems for \(\kappa\)-
and \(\zeta\)-\LSOP/
are very similar to the ones shown in Figure~\ref{fig:lbPP}.
\(\kappa\)-\LSOP/
illustrates that box constraints can be used to obtain \LSOP/ filters
with good convergence rates.  Nevertheless, larger box constraints
usually tend to result in worse rates of convergence, as compared to
an unconstrained filter.  While \(\kappa\)-\LSOP/
is still marginally better than the Elliptic filter, it yields rates
of convergence that are not nearly as good as \(\zeta\)-\LSOP/
or \(\eta\)-\LSOP/.

%% file: concl.tex
\section{Related Work} 
\label{sec:relatedWork} 


Contour based eigensolvers can be described as a family of methods
based on Cauchy's Residue Theorem.  The name ``contour based'' stems
from an integration of the matrix resolvent along a contour.  These
contour integrals are calculated via numerical quadrature which
inevitably results in linear system solves and reduced eigenvalue
problems through a Rayleigh-Ritz projection. Recently these methods
have gained popularity when used to solve the interior eigenvalue
problem. This popularity is in part due to the eigensolver's support for
non-Hermitian and generalized eigenproblems, and to an additional level
of parallelism compared to traditional iterative methods. One important
aspect of contour methods is that they typically tend to avoid
operations on long vectors, e.g, re-orthogonalization.

Two well-known contour based eigensolvers are the SS family of
solvers due to Sakurai and Sugiura~\cite{sakurai2003projection}, and
FEAST due to Polizzi~\cite{PhysRevB.79.115112}.  Early work by Sakurai
and Sugiura on contour based solvers resulted in the SS-Hankel method,
a non-iterative method based on complex moment matrices.  
A different approach is taken by Polizzi's FEAST, which is an
iterative contour based solver.  Underlying FEAST is a Rayleigh-Ritz
procedure where the contour integral is used to improve
convergence. Later work by Sakurai et.  al. resulted in a
Rayleigh-Ritz type method \cite{ikegami2010contour,Ikegami20101927}
and various block methods \cite{Imakura2016}.  Beyn proposed a similar
method in \cite{beyn2012integral}. Despite the rich variety of recent
works, contour based solvers are still a very active area of research
where most of the focus is on improving the performance and robustness
of the algorithms and their implementations. A recent comparison of
common contour based solvers is available in \cite{CompImakura2016}.

A key insight to the use of numerical quadrature for the contour
integration is the correspondence between contour methods and rational
matrix functions: one can reinterpret the numerical quadrature of the
matrix resolvent $(A-zI)^{-1}$ as a rational filter. From this point
of view contour methods can be considered as part of the richer
mathematical field of rational function approximation. The
correspondence between contour methods and rational filters was
discussed for FEAST in ~\cite{_feast_as_subsp_iterat_eigen}, for SS-H
in~\cite{Ikegami20101927}, and for SS-R in~\cite{ikegami2010contour}.
Early on Murakami proposed the use of classic rational filters from
signal processing
\cite{weko_67773_1,weko_69671_1,weko_28810_1,weko_28690_1,weko_18206_1}
in the ``filter diagonalization method''; 
for a discussion of this work we refer to
\cite{_comput_eigen_real_symmet_matric}.  For FEAST the Elliptical
filters were proposed in \cite{2014arXiv1407.8078G} to improve
load balancing and (in some cases) convergence behavior.  The rational
function point-of-view also enables the development of specialized
algorithms: for instance, for real-symmetric matrices it is possible
to avoid complex-arithmetic
entirely~\cite{_comput_eigen_real_symmet_matric}.

More recently the filter itself has been treated as a parameter that
can be designed via optimization methods.  Van
Barel~\cite{VanBarel2016346} proposed a non-linear Least-Squares
approach for non-Hermitian filters within the SS-H framework, while Xi
and Saad~\cite{saad16LeastSquares} described linear Least-Squares
optimized filters for the Hermitian FEAST solver. Since Van Barel's
approach is geared towards non-Hermitian eigenproblems, it is not
possible to make use of the conjugate or reflection symmetry. The
resulting degrees of freedom make for a more difficult optimization
problem. By requiring a parameter space search, his approach results
in filters that are not as robust as our \LSOP/.  Van
Barel's Least-Squares optimization is based on the discrete $\ell_2$
norm, not a function approximation approach.  Even in the 
case of Hermitian problems, optimizing the squared distances in
only a sample of points must be done with great care.  Choosing too
small a number of sample points can have undesirable effects: it is
possible to obtain poles near the real axis which are not detected by
the sample points. On the other hand having a large number of sample
points is very expensive.  Finally, Van Barel's approach does not
support constraints, as we present in Section~\ref{sec:constrOptimization}.

A FEAST related approach for linear Least-Squares optimized filters
\cite{saad16LeastSquares} focuses on the Hermitian
eigenproblem.
In this work, not the poles but only the coefficients of the rational
function are optimized. The result is a robust process that is much
easier to solve, at the cost of being less expressive.  An
optimization approach that does not optimize poles is limited by the
initial choice of them. As we have shown in this work, the
process of optimizing a \LSOP/ filter significantly moves the poles
inside the complex plane. On the opposite, fixing the poles imposes a
constraint on the optimization that is even larger than constraining
just their imaginary part as illustrated in
Section~\ref{sec:additional}. In practice, when optimizing solely the coefficients
of the rational approximation, the optimization algorithm behaves very
differently. For example, positive penalty parameters work much better in Xi
and Saad's approach, because real poles cannot occur.


\section{Conclusions}
\label{sec:conclusion}
In this work we illustrate how a weighted non-linear Least-Squares
optimization of rational filters provides a rich framework of
solutions that can be employed in contour based eigensolvers. In our
approach we optimize both the pole placement and the coefficients of
the rational filters, which leads to a non-convex problem. Because of
its non-convexity the solution of such problem has to be approached with care.
First we stabilize the
optimization process by explicitly formulating the filters to be
conjugation and parity invariant. In addition to stabilizing the
target function, such symmetries reduce the complexity of the
optimization process by a factor up to four. Due to the non-convexity
of the problem, convergence to an optimal solution is not easily
guaranteed. We show that a careful selection of the starting position and
an appropriate optimization algorithm, e.g. the Levenberg-Marquardt method,  lead to
solutions with systematically smaller residuals as the degree of the filter
 is increased.

 We show that using our optimized rational \LSOP/ filters can
 significantly improve the efficiency of contour based eigensolvers on
 a given problem. In particular we show that an optimized filter can
 improve the convergence of the subspace iteration for a large set of
 cases. In the specific instances of the standard Gauss and Elliptic
 filters we show-case the flexibility of our approach by providing
 \LSOP/ replacements. This flexibility is achieved via a skillful
 selection of Least-Squares weights and by the use of constrained
 optimization. We illustrate that optimization with box constraints
 can also usefully address the issue of rational filters with poles
 very close to the real axis, which lead to almost singular matrix
 resolvents. Such constrained optimization decreases the condition
 number of the resolvents, which positively influences the solution of
 the corresponding linear system solves when Krylov based methods are
 employed.

Significant effort went into the usability of the optimization
process. The entire approach is designed to be fire-and-forget,
without requiring many optimization processes for parameter space
exploration.  While we largely succeeded in our aim, there are still
corner cases where some parameter space exploration is required to
obtain usable filters. For instance, some trial-and-error is
required for constrained optimization. More robust
 optimization approaches, especially for constraints, is a
topic for future work. The ultimate goal is to provide filters that
adapt to a given specific problem. Such filters would be generated
on-the-fly and take advantage of spectral information, if cheaply
available. Not only do problem-specific filters promise better
convergence, but they can provide automatic load balancing between
multiple contour ``slices''. The present work is meant as a
significant step towards such a direction.

%% file: appendix.tex
\section{Definite integrals}
\label{sec:appendix}

For the convenience of the reader, we provide the definite integrals
required to compute the residual level function $F(\gamma, \omega)$
from Equation~\eqref{eq:reslev3}, the gradient $\nabla F$ from
Equations~\eqref{eq:gradFomega} and \eqref{eq:gradFomega}, and the
approximation of the Hessian, given in Equation~\eqref{eq:hessian}.
For the sake of generality we denote poles as $\omega_1$ and
$\omega_2$, without conjugation although some formulation require the
conjugation or sign change of $\omega_1$ or $\omega_2$.  Recall that
the functions $\WeightFct(t)$ and $h(t)$ are piecewise constant and determine
the integral boundaries, written as $a$ and $b$.

The calculation of $W_{k,\ell}$, $\pty{\cnj W}_{k,\ell}$, $X_{k,\ell}$, $\pty{\cnj X}_{k,\ell}$, $Y_{k,\ell}$, and $Z_{k,\ell}$ requires the following integral:
\begin{align*}
  \int_{a}^{b} \frac{ \ dt}{(t-\omega_1) (t-\omega_2)}
  &=& \frac{1}{\omega_1 - \omega_2} \left[\ln \frac{(b-\omega_1)}{(a-\omega_1)} +
    \ln \frac{(a-\omega_2)}{(b-\omega_2)}\right]
\end{align*}

For the 'diagonal' terms $W_{k,k}$ and $\pty{\cnj X}_{k,k}$ it holds than $\omega_1 = \omega_2$.
In this case we can re-write the expression without logarithms.
This integral is also required for $\nabla\!\cnj{\theta}_{k}$:
\[
\int_{a}^{b} \frac{ \ d t}{(t-\omega_1)^2} = \
\frac{b-a}{(b-\omega_1)(a-\omega_1)}
\]
Note that this is not the case for the $X$, $Z$, and $Y$, where the signs and/or conjugations of $\omega_1$ and $\omega_2$ differ.

For ${\theta}_k$ and $\pty{\cnj \theta}_k$ we require the following integral:
\[
\int_{a}^{b} \frac{\ d t}{t-\omega_1} = \ \ln \frac{(b-\omega_1)}{(a-\omega_1)}
\]
The calculation of $\nabla\!\cnj{W}_{k,\ell}$, $\nabla\!X_{k,\ell}$,
$\nabla\!\cnj{Y}_{k,\ell}$, and $\nabla\!Z_{k,\ell}$ requires the
integral:
\begin{align*}
  \int_{a}^{b} \frac{\ d t}{(t-\omega_1)^2(t-\omega_2)}
  = &\ \frac{b-a}{(\omega_1 - \omega_2)(b-\omega_1)(a-\omega_1)} - \frac{\ln \frac{(b - \omega_1)}{(a - \omega_1)}}{(\omega_1 - \omega_2)^2} +
  \frac{\ln \frac{(b - \omega_2)}{(a - \omega_2)}}{(\omega_1 - \omega_2)^2}
\end{align*}
The 'diagonal' term $\nabla\!\cnj{W}_{k,k}$ can again be expressed without logarithm:
\begin{equation*}
  \int_{a}^{b} \frac{dt}{(t - \omega_1)^3}
  =  \ (b-a) \left[\frac{b+a-2\omega_1} {(b-\omega_1)^2(a-\omega_1)^2}\right]
\end{equation*}

In the approximation of the Hessian the terms
$\nabla\nabla\!\cnj{W}_{i,j}$, $\cnj{\nabla}\nabla\!X_{i,j}$,
$\nabla\nabla\!\cnj{Y}_{i,j}$, and $\cnj{\nabla}\nabla\!Z_{i,j}$
require yet another integral. This one is given by:
\begin{align*}
  \int_{a}^{b} \frac{\ d t}{(t-\omega_1)^2 (t-\omega_2)^2}
  = & \  \frac{a-b}{(\omega_1-\omega_2)^2} \left[  \frac{1}{(a-\omega_1)(-b+\omega_1)} + \frac{1}{(a-\omega_2)(-b+\omega_2)} \right] \\
   + & \ \frac{2}{(\omega_1-\omega_2)^3} \left[ \ln \frac{a-\omega_1}{b-\omega_1} + \ln \frac{b-\omega_2}{a-\omega_2} \right]
\end{align*}
Here the 'diagonal' term is given by:
\begin{align*}
  \int_{a}^{b} \frac{\ dt}{(t-\omega_1)^4}
    = & \ \frac{1}{3} \left[ \frac{1}{(a-\omega_1)^3} + \frac{1}{(-b+\omega_1)^3}  \right]
\end{align*}

%% file: ssst.tex
In this short appendix we discuss a method to generate a large set of
intervals $\mathcal I$ from a given eigenproblem corresponding to an
Hermitian matrix $A$. The objective is to obtain a benchmark test set
for filtered subspace iteration, as a means to compare filters without
having to solve each problem separately.

Interior eigensolvers based on filtered subspace iteration are
sensitive to the spectrum of $A$ near the search interval $\mathcal
I$.  Thus, the choice of the search intervals is critical for testing
an interior eigensolver.  Users often select the search interval
motivated by some structure in the spectrum of $A$, such as the
HOMO-LUMO gap in Density Functional Theory methods.  Our aim is to
create a large set of intervals that represent a variety of real-world
use-cases, especially those based on the selection of intervals
influenced by the structure of the spectrum.  To this end, we propose
a method to obtain search intervals by exploiting the distribution of
the eigenvalues of $A$.  Instead of choosing the intervals directly,
we compute a set of endpoints $ E = \{ a_1, a_2, ... \} $, and then
form intervals $\mathcal I_{i,j} = [a_i,a_j]$ with $a_i, a_j \in E$
and $a_i < a_j$.

These endpoints should represent the neighborhood of an identifiable
spectral feature, such as a spectral gap or a cluster; we refer to
them as ``feature points''.  Given a function $\phi \in C^2$ that
approximates the eigenvalue density, we consider feature points to be
the real zeros of the first and second derivative of $\phi$.  A good
candidate for $\phi$ can be obtained, for example, via the Kernel
Polynomial Method (KPM) which constructs a polynomial of degree $M$
that approximates the spectral density.  The zeros of $\phi'$ are
stationary points; assuming that $\phi$ is a good approximation of the
eigenvalue density, a local maximum may indicate a high density of
eigenvalues or even a cluster. 
corresponds to a potential gap in the spectrum.  Inflection points of
$\phi$ may indicate a change in the increase or decrease of the
(approximated) eigenvalue density, signaling a relatively small
spectral gap.  By choosing a large $M \approx 45$ it is possible to
obtain thousands of intervals for a single matrix $A$.  The large
problem set can be used to obtain the subspace convergence rates and
the condition numbers of the linear system solves without solving each
eigenproblem in the benchmark set.  Such an approach is limited by the
availability of the entire spectrum of $A$, and thus is only
applicable for small to medium sized eigenproblems.

For instance, the subspace convergence rate (see Equation
\eqref{eq:convratio}) depends only on the filter value for the
eigenvalues.  If the entire spectrum of $A$ is available, we can obtain
the convergence rate by computing these values directly.
In the same fashion, each filtered interval results in a number of
shifted linear system solves of the form $A - Iz_i$ (see Equation
\eqref{eq:filter}).  When Krylov-based methods are used to solve these
systems, the overall performance depends on the condition number.
Since $A$ is Hermitian, the shifted matrix is still normal.  As a
result the singular values of $A-Iz_i$ are given by $\sigma_i =
|\lambda_i - z_i|$, where $\lambda_i$ is the \(i\)\textsuperscript{th}
eigenvalue of $A$.  By proceeding in this manner for all the poles
$z_i$, we can inexpensively obtain the condition number for all the
system solves. Usually, the solve with the highest condition number
dictates the overall performance, which is what we used to motivate
our analysis of box constraints in Section \ref{sec:boxconstraints}.

%% file: filters.tex
The filters presented in Section~\ref{sec:results} are provided here.
For a given filter function
\begin{equation*}
f(t,w,\gamma ) = \sum^q_{k=1}
\left[ \frac{\gamma_k}{t - w_k} + \frac{\cnj{\gamma}_k}{t - \cnj{w}_k}
  - \frac{\gamma_k}{t + w_k} - \frac{\cnj{\gamma}_k}{t + \cnj{w}_k} \right]
\end{equation*}
we provide only the poles in the "first quadrant" of the complex
plane---that is, the poles with positive real and imaginary
parts---and the corresponding coefficients \(\gamma\).\\

\newcolumntype{A}{>{\centering\arraybackslash}p{1.8cm}}%
\newcolumntype{B}{>{\centering\arraybackslash}p{1.7cm}}%
\newcolumntype{C}{>{\centering\arraybackslash}p{0.7cm}}%
\newcolumntype{D}{>{\centering\arraybackslash}p{1.4cm}}%
\newcolumntype{L}{>{\centering\arraybackslash}p{1.0cm}}%
\begin{center}
\begin{tabularx}{0.8\textwidth}{ X  D  B  B  D  D}
\toprule
$|t|\in$  & $[0, .95)$  & $[.95, 1.05)$ & $[1.05, 1.4)$   & $[1.4, 5)$  & $[5, \infty)$ \\
\midrule
  $\WeightFct_{\gamma\text{-\LSOP/}}(t)$ & 1 & .01 &  10 & 20 & 0  \\
\bottomrule
\end{tabularx}

\vspace{0.2cm}
\begin{tabularx}{0.85\textwidth}{ X  c c  c c c  }
\toprule
$|t|\in$  & $[0, .95)$  & $[.95, .995)$ & $[.995, 1.005)$ & $[1.005, 1.05)$ & $[1.05, 1.1)$\\
\midrule
$\WeightFct_{\eta\text{-\LSOP/}}(t)$ & 1 & 4 & 0.5 & 4 & 0.6\\
$\WeightFct_{\text{Box-\LSOP/}}(t)$ & 1 & 4 & 2 & 4 & 0.6\\
\bottomrule
\end{tabularx}

\vspace{0.2cm}
\begin{tabularx}{0.7\textwidth}{ X  c c  c c c  }
\toprule
 $|t|\in$  & $[1.1, 1.3)$ & $[1.3, 1.8)$ & $[1.8, 3)$ & $[3, \infty)$\\
\midrule
$\WeightFct_{\eta\text{-\LSOP/}}(t)$ (cont'd) & 1 & 0.3 & 0.1 & 0 \\
$\WeightFct_{\text{Box-\LSOP/}}(t)$ (cont'd)& 1 & 0.3 & 0.1 & 0 &\\
\bottomrule
\end{tabularx}

\captionof{table}{Weight functions for filters discussed in this paper.}
\end{center}


\begin{center}
\begin{tabular}{ll}
\toprule
Poles \(\omega\) & Coefficients \(\gamma\)\\
\midrule
\(0.999687712591797+0.0117367635577924i\)& \(0.006050012497458-0.000227036554136i\)\\
\(0.991596517222374+0.093208856178882i \)& \(0.021484299350510-0.003666847993474i\)\\
\(0.903848148311606+0.327740045699974i \)& \(0.055938387061383-0.032384567818443i\)\\
\(0.440319857798568+0.732970137475905i \)& \(0.054079510922005-0.122837701171251i\)\\
\bottomrule
\end{tabular}
\captionof{table}{\(\gamma\)-\LSOP/, a filter meant to replace the Gauss filter}
\end{center}

\begin{center}
\begin{tabular}{ll}
\toprule
Poles \(\omega\) & Coefficients \(\gamma\)\\
\midrule
\(0.999986323489133+0.002453510792541i \) & \(0.00110213725846833-7.98515806042 \cdot 10^{-6}i\)\\
\(0.999401600189637+0.024159213959740i \) & \(0.00768847889630669-0.000333793441122i         \)\\
\(0.983469964312691+0.160816338804574i \) & \(0.04592568742949140-0.008858565519222i          \)\\
\(0.628559997051189+0.718255201795431i \) & \(0.11139119375850727-0.147357486573937i          \)\\
\bottomrule
\end{tabular}
\captionof{table}{\(\eta\)-\LSOP/, a filter meant to replace the Elliptic filter. Obtained with a penalty parameter of \(c = -1.3  \cdot 10^{-6}\)}
\end{center}

\begin{center}
\begin{tabular}{ll}
\toprule
Poles \(\omega\) & Coefficients \(\gamma\)\\
\midrule
\( 0.999517437449349+0.0011346403206723i \) & \(0.000600799688893-0.0001380523176106i\) \\
\( 0.996122208058289+0.0169588203859498i \) & \(0.006148402177611-0.0012349951550185i\) \\
\( 0.971590779276380+0.1314326323772290i \) & \(0.039288214664051-0.0095779819369782i\) \\
\( 0.632009932807876+0.6589465004506030i \) & \(0.111997418650841-0.1454789177402232i\) \\
\bottomrule
\end{tabular}
\captionof{table}{\(\zeta\)-\LSOP/, a filter meant to replace the Elliptic filter that violates Guideline 1}
\end{center}

\begin{center}
\begin{tabular}{ll}
\toprule
Poles \(\omega\) & Coefficients \(\gamma\)\\
\midrule
\( 0.999997864241235+0.002199301304944i \) & \( 0.000920693720425-2.62225614322 \cdot 10^{-6}i  \) \\
\( 0.999817083735386+0.019255687759249i \) & \( 0.006309876577596-0.000106451402529i \)           \\
\( 0.993359173161426+0.135785991680408i \) & \( 0.040771709455088-0.003927802442013i \)           \\
\( 0.694622923894908+0.732441949783473i \) & \( 0.135841041422952-0.150283641573960i \)           \\
\bottomrule
\end{tabular}
\captionof{table}{\(\kappa\)-\LSOP/, a filter meant to replace the Elliptic filter with Krylov solvers that violates Guideline 1. Obtained with a box constraint of \texttt{lb} = 0.0022}
\end{center}